\documentclass[acmsmall, authorversion, nonacm]{acmart}

\usepackage{colortbl}
\usepackage{amsmath}
\usepackage{graphicx}
\usepackage{subcaption}
\usepackage{totpages}  % Ensure totpages package is included
\usepackage{mdframed}
\usepackage{titlesec}
\usepackage{enumitem}
\usepackage[normalem]{ulem}    % For strikethrough
\usepackage{xcolor}  % For coloring text

\AtBeginDocument{%
  }
\definecolor{darkgreen}{rgb}{0,0.5,0}
\definecolor{lightgray}{gray}{0.7}

\newcommand{\chicago}{Chicago}

\renewcommand{\paragraph}[1]{\vspace*{0.03in}\noindent\textbf{#1}}

\newcommand{\annotate}[1]{\textcolor{black}{#1}}

\definecolor{lightgray}{RGB}{220,220,220}

\NewDocumentCommand{\showhide}{mm}{%
    \ifstrequal{#1}{show}{%
        \begingroup
        \color{blue}#2%
        \endgroup
    }{%
    }%
}
\makeatother

\newcommand{\etal}{\textit{et. al.}}

\NewDocumentCommand{\shtext}{m}{%
  \showhide{hide}{#1}%
}

\long\def\showhide#1#2{%
  \ifstrequal{#1}{show}{%
    \begingroup
    \color{blue}#2%
    \endgroup
  }{%
  }%
}

\newmdenv[
  backgroundcolor=lightgray,
  linewidth=0pt,
  leftmargin=10pt,
  rightmargin=10pt,
  innerleftmargin=7.5pt,
  innerrightmargin=7.5pt,
  innertopmargin=5pt, 
  innerbottommargin=5pt,
  skipabove=\topsep,
  skipbelow=\topsep
]{takeawaybox}

\newcommand{\rev}[2]{#2}

\titlespacing*{\section}{0pt}{*1}{*1}  % Adjust the second and third arguments to reduce space before and after \section
\titlespacing*{\subsection}{0pt}{*0.5}{*0.5}  % Adjust for \subsection
\titlespacing*{\subsubsection}{0pt}{*0.5}{*0.5}  % Adjust for \subsubsection

\begin{document}

\title{Beyond Data Points: Regionalizing Crowdsourced Latency Measurements}

\author{Taveesh Sharma}
\affiliation{%
  \institution{University of Chicago}
  \city{Chicago}
  \country{USA}}
  \email{taveesh@uchicago.edu}  

\author{Paul Schmitt}
\affiliation{%
  \institution{University of Hawai\textquoteleft i, M\=anoa}
  \city{Honolulu}
  \country{USA}}
  \email{pschmitt@hawaii.edu}

\author{Francesco Bronzino}
\affiliation{%
 \institution{\'Ecole Normale Sup\'erieure de Lyon}
 \city{Lyon}
 \country{France}}
 \email{francesco.bronzino@ens-lyon.fr}

\author{Nick Feamster}
\affiliation{%
  \institution{University of Chicago}
  \city{Chicago}
  \country{USA}}
  \email{feamster@uchicago.edu}

\author{Nicole P. Marwell}
\affiliation{%
  \institution{University of Chicago}
  \city{Chicago}
  \country{USA}}
  \email{nmarwell@uchicago.edu}

\begin{abstract}
    
Despite significant investments in access network infrastructure, universal
access to high-quality Internet connectivity remains a challenge. Policymakers often rely
on large-scale, crowdsourced measurement datasets to assess the distribution
of access network performance across geographic areas. These decisions
typically rest on the assumption that Internet performance is uniformly
distributed within predefined social boundaries, such as zip codes, census
tracts, or neighborhood units. However, this assumption may not be valid for
two reasons: (1) crowdsourced measurements often exhibit non-uniform
sampling densities within geographic areas; and (2) predefined social
boundaries may not align with the actual boundaries of Internet
infrastructure.

In this paper, we present a spatial analysis on crowdsourced datasets for
    constructing \textit{stable} boundaries for sampling Internet performance.
    We hypothesize that greater stability in sampling boundaries will reflect
    the true nature of Internet performance disparities than misleading
    patterns observed as a result of data sampling variations. We apply and
    evaluate a series of statistical techniques to: (1) aggregate Internet
    performance over geographic regions; (2) overlay interpolated maps with
    various sampling unit choices; and (3) spatially cluster boundary units to
    identify contiguous areas with similar performance characteristics. We
    assess the effectiveness of the techniques we apply by comparing the similarity of
    the resulting boundaries for monthly samples drawn from the dataset. Our
    evaluation shows that the
    combination of techniques we apply achieves higher similarity compared
    to directly calculating central measures of network metrics over census tracts or neighborhood boundaries. These findings
    underscore the important role of spatial modeling in accurately assessing
    and optimizing the distribution of Internet performance, which can better
    inform policy, network operations, and long-term planning
    decisions.
\end{abstract}

  


\setcopyright{rightsretained}
\acmJournal{POMACS}
\acmYear{2024} \acmVolume{8} \acmNumber{3} \acmArticle{34}
\acmMonth{12}\acmDOI{10.1145/3700416}

\received{August 2024}
\received[revised]{September 2024}
\received[accepted]{October 2024}

\maketitle

\section{Introduction}\label{sec:intro}

Measuring the performance of Internet access networks is critical for
characterizing the quality of service that ISPs deliver to users
\cite{clark2021measurement} and for identifying discrepancies in Internet
performance in both urban and rural areas~\cite{jiang2023mobile,
sommers2012cell}. Over the past few decades, there have been significant
advancements in measuring access network performance, both in terms of novel
measurement infrastructure \cite{huang2008measuring, sundaresan2014bismark}
and analysis techniques \cite{sundaresan2011broadband,
sundaresan2017challenges, bischof2017characterizing,saxon2022we}. These
advances have primarily \rev{been}{~} focused on measuring the performance of a \textit{single} access link, using metrics such as throughput, latency, jitter, and
packet loss.  The Measurement Lab (M-Lab) \cite{mlab} and Ookla Speedtest
\cite{ookla} datasets are currently widely used for understanding Internet
performance of an access link \cite{broadband_mapping_coalition_2022}. Their
increasing prevalence has also enabled researchers to use these datasets to
ask a broader set of questions about Internet performance across an ISP or a
region~\cite{FCC2022, NYAG2020, BattleForTheNet2022,
PennStateMeasurementLab2019}. In particular, there has been significant recent
interest in understanding the distribution of Internet performance across
different geographies~\cite{lee2023analyzing, paul2023decoding,
sharma2023first, clark2021measurement}, especially for specific social and
policy-related inquiries. 

Yet, utilizing these crowdsourced measurements to characterize Internet
performance across a geographic region introduces new challenges, given the
nature of their \rev{measurements}{data}.  Most notably, crowdsourced datasets are
self-selected, both in time and in space. Both
the Ookla and M-Lab datasets generate crowdsourced “point" measurements from a
subset of Internet users across different geographies. These measurements,
often irregularly concentrated over space, are performed whenever a user
decides to run a speed test. Consequently, these data points reflect only a
small, non-uniform fragment of the overall user base and geographic area, posing significant
challenges for a comprehensive analysis. Yet, both policy and operational
decisions affecting a geographic region
rely on applying robust spatial analysis techniques to these small,
self-selected samples to make generalizations about Internet performance for
the entire resident population across that respective region. 

Spatial analysis can potentially transform these scattered data points into
cohesive insights, identifying patterns and trends that are not immediately
apparent. One significant challenge to this approach is identifying geographic
sampling boundaries for Internet performance and determining methods to
summarize these point measurements over space. Additionally, individual
measurements can be significantly noisy due to various factors such as testing
infrastructure, access media, and the client's hardware or software
platform~\cite{paul2022importance,macmillan2023comparative}. Spatial
de-noising and aggregation of these measurements is, therefore, critical for
drawing meaningful conclusions about network performance over specific
geographies. Analysis based on such de-noised datasets, on the other hand, can
ultimately help pinpoint areas \rev{}{truly} needing further infrastructure investments.

Prior work on applying spatial analysis to Internet measurements has taken a
different approach, treating spatial boundaries as given and applying
aggregation techniques within these pre-defined boundaries. For example,
previous work has attempted to characterize \rev{the}{} Internet performance over
conventional boundaries such as zip codes, census tracts, or neighborhood units~\cite{paul2021characterizing, saxon2022we,
sharma2022benchmarks,fiduccia2022deconstructing,jiang2023mobile}.  An
important conclusion from previous work is that there is often significant
variation in Internet performance over these
boundaries~\cite{sharma2022benchmarks}; previous work has suggested that such
regions need additional attention or policy intervention. However, these
approaches suffer from a few important limitations. First, the use of
aggregate measures such as mean, median and inter-quantile range (IQR)
\cite{paul2021characterizing,sharma2022benchmarks} on point measurements may
lead to inaccurate conclusions across a region, particularly when the region
exhibits high variability or is poorly sampled.  For example, measurements in
pre-defined regions are often clustered in some portions of the space and
dispersed in others \cite{jiang2023mobile}. Thus, any aggregate measures
calculated over irregularly clustered data may overrepresent densely sampled
areas. 

Second, to our knowledge, no previous work has assessed the accuracy of
previous techniques~\cite{fiduccia2022deconstructing, jiang2023mobile} in
summarizing Internet performance over a pre-defined geography. This may
prevent regulators and ISPs from adopting the most effective aggregation
techniques for their analyses. Finally, correlating Internet performance with
population measures such as median income and population density
\cite{fiduccia2022deconstructing,saxon2022we,paul2021characterizing} using
existing social boundaries may \rev{not}{} be inappropriate due to imperfect alignment
with infrastructure boundaries. A single social boundary may encompass
multiple areas with heterogeneous Internet infrastructure, potentially leading
to misleading correlations. Simply put, there is no reason to expect that
Internet performance should be spatially aggregated along human-defined
boundaries that have nothing to do with the Internet itself. This paper is the
first to explore and evaluate alternate spatial clustering approaches that
more accurately reflect relationships in the underlying Internet measurement
data.

Our work addresses these concerns by applying a new combination of
statistical techniques to aggregate point measurements over a geography and
discover \textit{stable} sampling boundaries, that is, boundaries that show
less variability when subject to variations in the underlying data. We
hypothesize that optimizing for stability will make it easier to compare data
across different regions, time periods, and longitudinal studies. With
consistent boundaries, we expect reduced influence of variability in the
underlying data, which is likely to reflect true differences rather than
artifacts of boundary shifts. This consistency is critical
for accurate spatial analysis, as stable geographic boundaries will enable
researchers, operators, policymakers, and others to track changes
over time, compare different geographic areas, and conduct longitudinal
studies with greater confidence in outcomes.

The solution we develop comprises three steps. We first use and compare prior
techniques to interpolate Internet performance to synthetic, out-of-sample
locations for areas that are otherwise unsampled in crowdsourced datasets.
Second, we use this capability to summarize latency within small, polygonal
tessellations of varying resolutions, census tracts, as well as neighborhood
boundaries within a large US city. Finally, we cluster these smaller units to
discover the edges of sampling boundaries.  We focus on latency because this
metric is increasingly critical to user quality of experience for
latency-sensitive applications, such as Web browsing, interactive video, and
gaming; latency is also an important differentiator between conventional
fixed-line ISPs and emerging fixed 5G providers and is thus an important
metric to study. Although we focus on this single metric for this paper, as
our focus is on applying the spatial analysis techniques themselves, we expect
that the techniques that we develop are broadly applicable across other
metrics. 

To evaluate the quality of the
resulting clusters, we measure the similarity between boundaries using the
Adjusted Rand Index (ARI) \cite{steinley2004properties} for monthly samples
drawn from the interpolated dataset. We show that these techniques achieve a
median pairwise ARI score of 0.59 (on a scale of -1 to 1), which provides a
0.39 gain over computing raw averages for census tract boundaries. An ARI
score of 0.59 indicates a moderate to strong agreement between the clustering
results for independent monthly fits, demonstrating that the clusterings capture
significant spatial structure in the data. Our work makes the following
contributions:

\begin{itemize}[left=10pt, topsep=0pt, partopsep=0pt, parsep=0pt]
    \item We develop an end-to-end analysis pipeline to construct \rev{}{stable} measurement-driven boundaries for sampling Internet performance over a large US city\rev{to achieve stable sampling boundaries}{~}. Our boundaries show consistency across monthly samples drawn from the same dataset, up to an Adjusted Rand Index of 0.59.
    \item We demonstrate how and when ISPs and regulators can use our techniques to identify areas with similar latency characteristics from a given sample of measurement data. For instance, we show that using the $95^{th}$ percentile of latency for spatial aggregation yields more stable clusters than using the $10^{th}$, when homogenous clusters covering small geographic areas are desirable.
    \item We find that boundaries constructed from 17-month-long, ISP-specific data samples do not show significant similarity between ISPs. This suggests that \rev{a one-size-fits-all measurement data sampling approach may not be appropriate for all ISPs.}{the FCC should consider releasing ISP-specific representations of Internet performance for greater transparency.}
    \item While network operators may deploy their own measurement infrastructures, our approach offers significant advantages by utilizing crowdsourced data, allowing coverage from multiple real-world vantage points in a cost-effective manner. We release our source code for constructing these boundaries, enabling the research community, policymakers, and ISPs to use it in their analyses \cite{artifacts}.
\end{itemize}
  \section{Background}\label{sec:background}

  \shtext{(1) Two approaches to discovering boundaries -- use coverage data from ISPs, or construct statistical boundaries. We chose the second approach.}

  We describe processes for summarizing Internet performance within a
  geography; then, we discuss spatial interpolation and clustering techniques
  for identifying boundaries for data across a geography that could ultimately
  be applied to Internet performance measurements.

\subsection{Sampling Internet Performance in a Region: Two Approaches}

Discovering boundaries for sampling Internet performance in a region can be
done with two possible approaches: Targeted data collection within a region,
and statistical interpolation of existing, crowdsourced data. 

The first approach involves collecting data from ISPs and aligning sampling
boundaries with coverage maps that are regularly updated by the FCC \cite{NBM}. The accuracy of these maps has
recently come under scrutiny~\cite{natlawreview}, which, in the United States,
has given rise to the Broadband Equity Access and Deployment (BEAD) program's
Challenge process \cite{marques2024we}. The BEAD Challenge process is designed
to allocate federal funding for broadband infrastructure projects across the
United States, particularly in underserved regions. To enhance broadband
availability maps across the country, participants in this process are
required to submit accurate coverage data by running local measurement
campaigns. The challenge process is ongoing
\cite{internetforall2024challenge}, with states and territories submitting
their data to the National Telecommunications and Information Administration
(NTIA) for review. Creating accurate coverage maps is a
future objective that involves extensive regulatory considerations. 

An alternative approach is to analyze the statistical distribution of existing
crowdsourced measurement data from speed test providers such as Ookla or
M-Lab. A key challenge to this approach is the under-representation of areas
where users are less likely to conduct speed tests. It is thus important to
apply post-collection analysis techniques that accurately characterize
Internet performance in sparsely sampled areas. 

\subsection{Spatial Interpolation and Clustering of Crowdsourced Measurements}

In this work, we adopt the second approach. To address the challenge of data
sparsity in crowdsourced measurements, we apply and evaluate spatial
interpolation techniques in the context of Internet measurement data. We then
explore the use of a spatial clustering technique to identify geographic
boundaries for sampling Internet performance, given this interpolated data.
In this section, we provide an overview of relevant spatial statistics
literature.

\subsubsection{Spatial Interpolation.} \label{sec:interpolation}

There are two types of interpolation techniques: deterministic and stochastic.
Deterministic techniques make mathematical assumptions about the spatial
process to predict the target variable without incorporating randomness in the
process. Examples of deterministic techniques include Inverse Distance
Weighting (IDW) \cite{shepard1968two}, LOESS \cite{Cappellari2013b}, and
Self-tuning Bandwidth in Kernel Regression (STBKR) \cite{jiang2023mobile}.
While Kriging \cite{cressie1988spatial} is often considered deterministic in
application, it is based on a stochastic model and can provide uncertainty
estimates, making it somewhat of a hybrid technique.

Stochastic techniques, on the other hand, incorporate randomness and
statistical properties of the spatial data to yield predictions along with
uncertainty estimates at each location. These techniques are more appropriate
when there is strong spatial dependence in the underlying data. Examples of
stochastic techniques include Gaussian processes \cite{aldworth1999sampling},
Random Forests \cite{sekulic2020random}, and Neural networks
\cite{rigol2001artificial}. Gaussian processes model spatial data as a
collection of random variables, where the covariance between any two variables
is a function of the distance between them. Random forests are stochastic due
to the randomization involved in their construction procedure, while the
stochasticity in neural networks stems from weight initialization and gradient
descent process.

In the context of crowdsourced network data, we argue that the noise
introduced by factors such as Wi-Fi
\cite{sharma2023measuringprevalencewifibottlenecks} and access equipment
\cite{sundaresan2011broadband} may weaken spatial auto-correlation between
neighboring measurements. This can make deterministic techniques more
suitable, as they do not rely on the statistical properties of the data as
much as stochastic techniques. In this paper, we thus use three deterministic
techniques---IDW, LOESS, and STBKR---for comparison and \rev{extension}{design an analysis pipeline that can be integrated with any spatial interpolation method}. 
Exploration of stochastic techniques for interpolating Internet measurement
data is a ripe avenue for future work.

\subsubsection{Spatial Clustering \& Regionalization.} \label{subsec:clustering-bg}

\shtext{(1) Regionalization = Spatial clustering + contiguity constraints (2) Why SKATER -- graph-based methodology.}

Spatial clustering involves the process of grouping similar data points
based on their spatial proximity, or sometimes another attribute of
interest. Common spatial clustering algorithms include K-Means, DBSCAN, and
Hierarchical Clustering. The output of applying these algorithms to spatial
data is a set of clusters, which may or may not be contiguous in space. A
specific form of spatial clustering is regionalization, also known as
spatially constrained clustering. Clusters formed using regionalization are
contiguous in space. Common regionalization algorithms include the Automatic
Zoning Procedure (AZP) \cite{anselin2018spatial}, the Max-P algorithm
\cite{duque2012max}, and Spatial `K'luster Analysis by Tree Edge Removal
(SKATER) \cite{assunccao2006efficient}. In this work, we consider the use of
regionalization in identifying areas with similar Internet performance in a
citywide geography. 

Our analysis is limited to regionalization techniques because of our prior
assumptions about Internet infrastructure. Internet infrastructure is often
laid out hierarchically in contiguous regions, with local networks aggregating
into regional, and ultimately into core networks \cite{ge2001hierarchical}.
SKATER uses a tree-based methodology, under which it tries to hierarchically
merge similar spatial units. Depending on the policy or other objectives, the
\textit{number of clusters} parameter in SKATER can be adjusted to identify
how local clusters are merged into regional or city-level clusters. This
characteristic makes SKATER a good fit for analyzing Internet performance
data.

\section{Method}\label{sec:methods}
\shtext{(1) Chose Ookla because it is large and has accurate geolocations.}

This section describes our analysis method. We first analyze latency
measurements from the Ookla dataset to discover statistical boundaries for
sampling Internet performance in a large US city. We then describe the data
preprocessing steps that we applied to the initial sample of Ookla measurements
to arrive at the dataset that we use for our analysis in this paper. This
preprocessing ensures that our interpolation and clustering analyses are least
affected by noise originating from a variety of factors that could distort the
sample, such as VPN connections (which can artificially inflate latency) and
inaccurate geolocation (which can create outliers that do not correspond to
spatial properties in the dataset). Given the dimensionality of the dataset,
there can be considerable variations in sample selection methods for
interpolation and clustering. Our approach is designed to ensure that our
results are reasonably robust to these variations. Finally, we describe our
analysis pipeline, which involves spatial interpolation to construct a uniform
surface model of latency values, followed by regionalization to identify
contiguous regions with similar latency characteristics.

\subsection{Scope of Analysis}

\paragraph{Dataset:} \footnote{\rev{}{Access to the Ookla dataset used in this paper can be obtained from \href{https://www.ookla.com/speedtest-intelligence.}{https://www.ookla.com/speedtest-intelligence}}} We use a proprietary Ookla dataset for our analysis as
Ookla provides the largest crowdsourced measurement dataset for access network
performance in the present day. As opposed to the M-Lab dataset, Ookla
provides access to a greater number of point measurements with high quality
geolocations, which is crucial for spatial analysis. Ookla uses a combination
of GPS and IP geolocation to triangulate a user. We found a lack of
availability of an accuracy measure for the geolocated measurements, which
renders these measurements unreliable for a high-resolution spatial analysis.
So, we choose to focus on GPS geolocated measurements. We analyze this dataset
for fixed line ISPs because these ISPs are likely to provide a more stable and
consistent service quality compared to mobile ISPs, which is desirable for
spatial analysis. Finally, we conduct our analyses for the city of \chicago{}
because (1) it provides the second-largest overall sample size at a city
level, and (2) it is a city with a well documented history of measurement
sampling bias across its subdivisions \cite{saxon2022we,sharma2022benchmarks}. 

\paragraph{Performance metric:} We use latency for spatial analysis due to its
strong correlation with geographic distance and its effectiveness as a proxy
for end-user quality of experience (QoE), compared to metrics like throughput
or packet loss \cite{bitag_latency_explained}. Higher latency often results in
increased buffering for real-time applications such as video streaming,
conferencing, and online gaming. Additionally, latency allows for comparisons
across different access technologies like DSL, Fiber, and Cable. In contrast,
throughput can be influenced by subscription tiers, local bottlenecks, and
server capacity, making it less suitable for spatial analysis. Packet loss,
being more sporadic and prone to getting influenced by transient issues like network
congestion, may not exhibit clear spatial patterns. Ultimately, analyzing
latency can help network operators and regulators pinpoint regions with poor
user satisfaction and guide targeted policy interventions to enhance Internet
quality.

\subsection{Data Preprocessing} \label{subsec:data-cleaning}

\begin{table}[t!]
    \centering
    \begin{tabular}{lr}
        \toprule
        \textbf{Preprocessing Step} & \textbf{Number of Samples Retained} \\
        \midrule
        Measurements for Chicago & 5,924,004 \\
        Non-VPN Measurements & 5,266,797 \\
        Auto-selected Server Measurements & 5,005,881 \\
        GPS-only Measurements & 891,431 \\
        \bottomrule
    \end{tabular}
    \caption{Summary of the filtering steps applied to the initial sample of Ookla measurements.}
    \label{tab:filtering-steps}
\end{table}

% The Ookla dataset contains a large repository of crowdsourced measurements. These measurements are collected using the Speedtest application \cite{ookla}, which is available across a wide range of users, platforms and devices. Given the large scale of this dataset, it is crucial to apply a series of filtering steps to narrow down our focus on a subset of measurements that are likely to provide insights into the geographic variations in latency, without being affected by noise. In this section, we thus describe a series of filtering steps applied to our initial sample of Ookla measurements. These steps are different from spatial interpolation/de-noising as they are applied to the raw data before performing any spatial analysis. Interpolation, on the other hand, allowed us to capture any spatial trends in the data over the citywide geography.

From a vast US-wide dataset, we focus on measurements that originated from
Chicago because this city provided a large sample size, has a rich set of
demographics, and there is evidence for considerable sampling bias across its
subdivisions. Next, we exclude measurements that are conducted over a VPN
connection. We do so to ensure that only the measurements conducted over the
user's home network are considered. VPNs can introduce additional latency and
may not be representative of the user's actual experience. Next, we filter out
measurements conducted against servers that are not auto-selected by Ookla to
conduct the speed test. We exclude these measurements to ensure that the
latency being analyzed is representative of typical latency experienced by
user-facing applications. 

\begin{table}
    \begin{tabular}{lr}
    \toprule
    \textbf{Descriptive} & \textbf{Value} \\
    \midrule
    Measurement Duration             & Jan 2022 -- Jun 2023   \\ 
    \# Measurements                 & 891,431               \\
    \# distinct ISPs                 & 799                    \\
    \# distinct vantage points (VP) & 133,427                \\ 
    Median \# samples per VP        & 2                     \\ 
    \# target servers              & 909                    \\ 
    \rev{}{\# hexagonal cells}               & \rev{}{899}              \\
    \rev{}{\# cells with lower than average sample size (sparse cells)}         & \rev{}{643}              \\
    \rev{}{Average \# measurements per cell per month}         & \rev{}{66.15}              \\
    \bottomrule
    \end{tabular}
    \caption{Basic descriptives of the final sample of Ookla measurements.}
    \label{tab:descriptives}
\end{table}

Most Content Delivery Networks (CDNs) tend to deploy their content caches
close to end users \cite{hasan2014trade} for achieving low latency. Ookla
defaults to nearby servers based on ping results for multiple servers,
ensuring that the selected server is the closest to the user. Using an
auto-selected server thus ensures that our analysis is not biased by the
user's choice of a distant server. Finally, we exclude all IP geolocated
measurements due to their lack of reliability for high resolution spatial
analysis. GeoIP measurements are expected to yield high location errors, which
could skew our findings. Though GPS is also prone to errors, we found in our
sample that a large proportion of GPS locations (87.4\%) were within a
460-meter radius of the true location. 460 meters is the size of a
resolution-8 hexagon in the H3 tessellation system \cite{uber_h3}, which we
use for our analysis in Section~\ref{sec:regionalization}. As an additional
consideration, we analyze the age of GPS locations in the final sample.
Across our 17-month sample, we found the $95^{th}$ percentile location age to
be about 63 seconds across all measurements, indicating that most locations
are reasonably recent. 

Table~\ref{tab:filtering-steps} summarizes the filtering steps used to build our analytic sample. Most reduction in the dataset size occurs at the initial filtering steps, which is expected because we are focusing on a specific geography. We believe that the final sample is representative of the population of users for multiple ISPs in a large US city, and is thus suitable for our analysis. We compare the quality of boundaries between ISPs in Section~\ref{sec:sample-effects}. Basic descriptive statistics for the final sample are summarized in Table~\ref{tab:descriptives}.

\subsection{ Analysis Approach}

\begin{figure*}[t]
    \centering
    \includegraphics[width=1.0\linewidth]{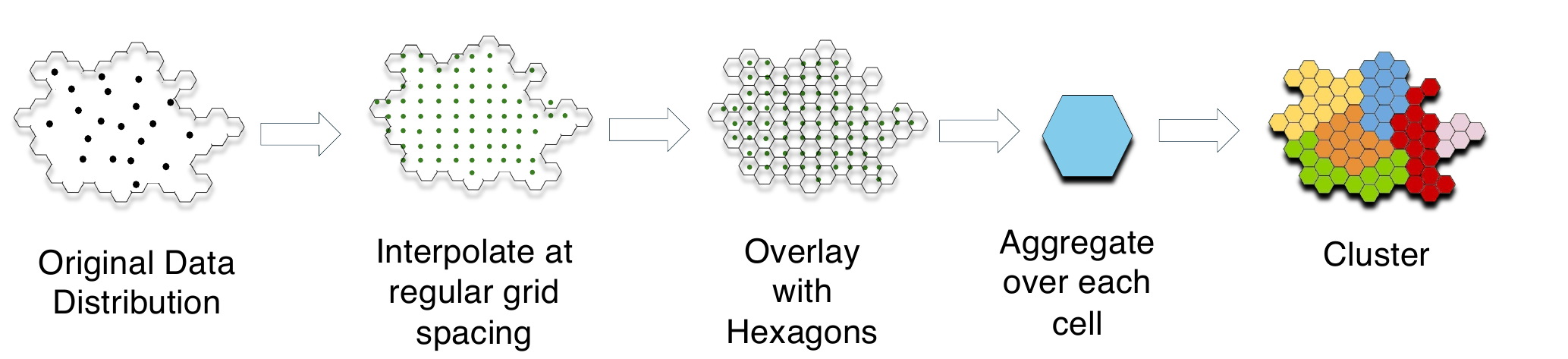}
    \caption{Overview of our analysis pipeline. \annotate{First, we construct an interpolated map of the region. Then, we use this map to perform spatial clustering.}}
    \label{fig:pipeline}
\end{figure*}

\shtext{(1) Overview. (2) Key assumptions. (3.1) Interpolation -- Choice of resolution (3.2) Interpolation Workflow and Evaluation metric (4) Regionalization (4.1) Parameter selection (4.2) Comparison with social boundaries. (4.3) Impact of interpolation.}
\shtext{(1) Overview.}

Given the unreliability of current FCC coverage maps, our analysis aims to establish statistical sampling boundaries for latency in Chicago. With latency measurements likely to be unevenly distributed across the geography, our first step is to develop a uniform surface model of latency values. This model allows us to estimate latency at unsampled locations and ensures that our conclusions are not biased by uneven sampling densities. We achieve this by applying spatial interpolation techniques.

We first evaluate the performance of three interpolation methods by predicting latency at in-sample locations. Then, using one of the methods, we predict latency at regularly spaced points on the map, which we further overlay with different boundary unit choices. Finally, we use the SKATER regionalization algorithm, which preserves spatial contiguity, to identify regions with similar latency characteristics. Figure~\ref{fig:pipeline} summarizes our pipeline and the analysis goals. We describe key aspects of this pipeline in detail below.

\paragraph{Interpolation at regular grid spacing.} We used an 80-20 split of the dataset to evaluate the chosen techniques across the city's geography. Due to the dataset's large size, we fit each model on monthly splits, resulting in 17 different fits per interpolation method parameter choice. Our evaluation ensures that all measurements from a single user appear exclusively in either the training set or the test set to prevent overfitting. Including a user's measurements in both sets could lead models to learn the user's latency distribution rather than the underlying spatial patterns. By treating each user independently, we assess how well the models generalize to unsampled users at a given location. Upon selecting a suitable interpolation method, we then predict latency at regularly spaced points across the city to construct a uniform surface model of latency. The grid points for this out-of-sample interpolation step are chosen to be spread across the city at a regular spacing of 50 meters. This choice is made to ensure that the interpolated map was evenly spaced, granular and smooth, which allows an unbiased calculation of latency aggregates such as averages and percentiles. \rev{}{We implement our interpolation workflow using standard Python libraries such as Scikit-learn \cite{scikit-learn} and Geopandas \cite{geopandas}, but tools such as ArcGIS \cite{arcgis} or QGIS \cite{qgis} can also be adopted for this step. The choice of Python libraries is motivated by their open-source nature, flexibility, and ease of integration with typical data science workflows.}

\paragraph{Overlay with Hexagons.} This process involves creating a tessellation of regular hexagons to comprehensively cover the entire study area of Chicago. Hexagons are chosen because they have the highest perimeter-to-area ratio among regular polygons, which allows them to tessellate the map with minimal overlap. The Federal Communication Commission (FCC) commonly uses the H3 tessellation system \cite{uber_h3} to map broadband availability across the United States. H3 is a hierarchical geospatial indexing system widely used in real-time applications such as taxi demand forecasting and urban planning. H3 hexagons can be constructed at 16 different resolutions ranging from 0 to 15, with a higher resolution representing hexagons of smaller edge lengths. The FCC uses a resolution of 8 for their broadband availability maps. To ensure compatibility with the FCC maps, we choose to use the same resolution for our analysis.

\paragraph{Aggregation over each cell.} Upon overlaying the interpolated points with regular hexagonal cells, we select a suitable clustering criterion for SKATER. We thus experiment with different aggregation choices such as mean, percentiles, standard deviation and other metrics used in prior work. The sensitivity of our approach to these metrics is discussed in Section~\ref{sec:agg-metric}. Ultimately, we apply and evaluate SKATER to perform regionalization on the aggregated cells to identify regions with similar latency characteristics.

\paragraph{Spatial Clustering.} Stable regional clusterings across time are expected to be relevant for effective policy-making, particularly in the context of funding programs like the BEAD initiative. BEAD funding aims to address disparities in broadband access by targeting resources to underserved areas. Consistent regional boundaries ensure that these investments are directed efficiently and equitably. Our approach, therefore, should result in clusters that identify consistently underserved regions regardless of the choice of time interval for drawing the data samples. We thus evaluate our regionalization results from a lens of stability across time. To measure stability, we use the Adjusted Rand Index (ARI) \cite{steinley2004properties} due to its ability to preserve the relative ordering of the clusters. ARI considers all pairs of hexagons and counts the number of pairs that are assigned to the same or different clusters between two clusterings. Then, it calculates the probability of agreement between the two clusterings, and compares it to the expected agreement under random assignment. The ARI score ranges between -1 and 1, with 1 indicating perfect agreement between the clusterings, 0 indicating random assignment, and -1 indicating complete disagreement. We use the ARI score to compare the clusterings obtained for different choices of SKATER parameters. For each parameter combination, we calculate the median ARI score between each pair of monthly clusterings. Median is chosen for this aggregation because we noticed multimodal behavior in the distribution of pairwise ARI scores over our parameter choices.

\section{Interpolation}\label{sec:comparison}

In this section, we compare three deterministic interpolation
techniques---Inverse Distance Weighting (IDW), Locally Estimated Scatterplot
Smoothing (LOESS), and Self-tuning Bandwidth in Kernel Regression (STBKR)---to
estimate latency at unsampled locations in the Ookla dataset. We evaluate the
precision and reliability of these techniques, and discuss the implications of
our findings.

% % Please add the following required packages to your document preamble:
% % \usepackage{graphicx}
% % \begin{table*}[t]
% %     \centering
% %     \begin{tabular}{|p{3cm}|p{7cm}|p{5cm}|}
% %     \hline
% %     \textbf{Interpolation Method} & \textbf{Description} & \textbf{Parameters} \\
% %     \hline
% %     Inverse Distance Weighting (IDW) & Computes weighted average of nearby measurements in proportion to their relative distance from an unsampled location. & $p$ (Impact of distance on weights) \\
% %     \hline
% %     Locally Estimated Scatterplot Smoothing (LOESS) & Fits local regression lines to de-noise latency across space. & $span$ (Proportion of data points for regression) \\
% %     \hline
% %     Self-tuning Bandwidth in Kernel Regression (STBKR) & Computes weighted averages of nearby measurements as estimates for unsampled locations using a Gaussian Kernel that models point densities. & $c$ (Controls bandwidth of the kernel), $k$ (Number of nearest neighbors) \\
% %     \hline
% %     \end{tabular}
% %     \caption{A summary of chosen interpolation methods and their parameters.}
% %     \label{tab:interpolation-methods}
% % \end{table*}

\begin{table*}[t]
    \centering
    \begin{tabular}{p{3cm}p{5cm}p{5cm}}
    \toprule
    \textbf{Interpolation Method} & \textbf{Description} & \textbf{Parameters} \\
    \midrule
    \rowcolor{lightgray} Inverse Distance Weighting (IDW) & Computes weighted average of nearby measurements. Uses the distance from an unsampled location for weight calculations. & $p$ (Impact of distance on weights) \\
    % \hline
    \addlinespace
    Locally Estimated Scatterplot Smoothing (LOESS) & Fits local regression lines to de-noise latency across space. Uses lat-long values directly for regression. & $span$ (Proportion of data points \rev{}{used} for regression) \\
    % \hline
    \addlinespace
    \rowcolor{lightgray} Self-tuning Bandwidth in Kernel Regression (STBKR) & Computes weighted averages of nearby measurements as estimates for unsampled locations. Uses a Gaussian Kernel to model point densities. & $c$ (Controls bandwidth of the kernel), $k$ (Number of nearest neighbors) \\
    \bottomrule
    \end{tabular}
    \caption{A summary of chosen interpolation methods and their parameters.}
    \label{tab:interpolation-methods}
\end{table*}

% In this section, we formally introduce the interpolation problem and the chosen techniques -- LOESS, STBKR and IDW. Finally, we present an empirical comparison of the three techniques in estimating latencies at unsampled locations in the Ookla dataset.

\subsection{Problem Formulation} \label{subsec:problem-formulation}

Assume that we are given \(n\) observed locations with latency values \(z_i\) at locations \((x_i, y_i)\), \(i = 1, 2, \ldots, n\), and we are interested in estimating the latency \(z\) at an unmeasured location \((x, y)\). Let \(Z(x, y)\) denote the latency value at location \((x, y)\) for a specific month. We are interested in obtaining an estimate \(\hat{Z}(x, y)\) for \(Z(x, y)\) at \((x, y)\). 
\subsection{Techniques}

For the above problem formulation, we consider three interpolation techniques: Inverse Distance Weighting (IDW), Locally Estimated Scatterplot Smoothing (LOESS), and Self-tuning Bandwidth in Kernel Regression (STBKR). We summarize these techniques in Table \ref{tab:interpolation-methods}.

\paragraph{Inverse Distance Weighting (IDW).} IDW assigns weights to each nearby data point based on its distance from an unsampled location. It uses these weights to calculate a linear combination of nearby values as an estimate of the target metric at an unsampled location. The relationship between the similarity of nearby data points and their distance is assumed to be inverse in nature. The IDW estimate \(\hat{Z}(x, y)\) at location \((x, y)\) is given by:

\[
  \hat{Z}(x, y) = \frac{\sum_{i=1}^{n} \frac{z_i}{d_i^p}}{\sum_{i=1}^{n} \frac{1}{d_i^p}}
\]

where \(d_i\) is the Euclidean distance between the target location \((x, y)\) and the \(i^{th}\) data point \((x_i, y_i)\), and \(p \geq 1\) is a parameter used to control the influence of nearby points. A higher value of \(p\) indicates a greater influence.

\paragraph{Locally Estimated Scatterplot Smoothing (LOESS).} LOESS \cite{Cappellari2013b} is a non-parametric regression technique that fits a smooth curve to a scatterplot of data points. By fitting a set of local polynomials to the spatial data, it smoothes any discontinuities and effectively captures the underlying spatial patterns. LOESS uses a smoothing parameter $\alpha$, commonly known as the span, to control the extent of smoothing. It assigns weights to the nearby data points $(x_i, y_i)$ depending on their distance from an unsampled location $(x, y)$ using a Tri-cube Kernel. The weights, $w\{(x, y), (x_i, y_i)\}$ are given by:

\[
\begin{cases}
  \left(1 - \left(\frac{\|(x_i, y_i) - (x, y)\|}{h}\right)^3\right)^3 & \text{if } \|(x_i, y_i) - (x, y)\| \leq h, \\
  0 & \text{otherwise}
\end{cases}
\]

% \begin{figure*}[t]
%     \centering
%     \begin{subfigure}[b]{0.32\linewidth}
%         \includegraphics[width=\textwidth]{figures/sec4/rmse_resolution_android.pdf}
%         \caption{Android Errors}
%         \label{fig:overall-android}
%     \end{subfigure}
%     ~
%     \begin{subfigure}[b]{0.32\linewidth}
%         \includegraphics[width=\textwidth]{figures/sec4/rmse_resolution_ios.pdf}
%         \caption{iOS Errors}
%         \label{fig:overall-ios}
%     \end{subfigure}
%     ~
%     \begin{subfigure}[b]{0.32\linewidth}
%         \includegraphics[width=\textwidth]{figures/sec4/method_precision.pdf}
%         \caption{Drop in RMSE range versus resolution}
%         \label{fig:precision-drop}
%     \end{subfigure}
%     \caption{Distribution of RMSE values for LOESS, STBKR and IDW across different resolutions for Android and iOS users. \annotate{The range of RMSE values decreases as resolution increases.}}
%     \label{fig:rmse-resolution}
% \end{figure*}

The bandwidth of the Kernel $h$ is set in such a way that approximately $\alpha \times n$ neighbors are included in each local regression, where $n$ is the total number of data points. $\|(x_i, y_i) - (x, y)\|$ denotes the Euclidean distance between a sampled and an unsampled location. The final estimate $\hat{Z}(x, y)$ at location $(x, y)$ is given by $\hat{Z}(x, y) = \hat{\beta_0} + \hat{\beta_1}x + \hat{\beta_2}y$. The coefficients $\hat{\beta}(x, y) = \{ \hat{\beta_0}, \hat{\beta_1}, \hat{\beta_2} \}$ are determined by minimizing the weighted sum of squared residuals, akin to traditional regression methods:

\[
\hat{\beta}(x, y) = \arg\min_{\beta} \sum_{i=1}^{n} w\{(x, y), (x_i, y_i)\} \{z_i - \hat{Z}(x_i, y_i)\}^2
\]

It is worth noting that the above formulation uses a linear polynomial. It is possible to use higher-order polynomials to fit the data, though this may lead to overfitting. In our work, we restrict the scope to linear polynomials due to their low complexity and high interpretability.

\paragraph{Self-tuning Bandwidth in Kernel Regression (STBKR).} The STBKR technique proposed in Jiang \etal~\cite{jiang2023mobile} uses a Gaussian Kernel regression method to estimate mobile Internet quality. Their approach allows the bandwidth of the Kernel to be tuned automatically, depending on the density of measurements in the local neighborhood of an unsampled location. The STBKR estimate of \(\hat{Z}(x, y)\) at location \((x, y)\) is given by:

\[
  \hat{Z}(x, y) = \frac{\sum_{i=1}^{n} K_{h(x, y)}\left(\| (x_i, y_i) - (x, y) \|\right) z_i}{\sum_{i=1}^{n} K_{h(x, y)}\left(\|(x_i, y_i) - (x, y) \| \right)}
\]

where \(K_{h(x, y)}\) is the Gaussian Kernel function with bandwidth \(h\), given by $K(u) = \frac{1}{\sqrt{2\pi}h(x, y)} e^{-\frac{u^2}{2h^2(x, y)}}$. The use of a Gaussian Kernel provides a mechanism for decaying the influence of data points as their distance from the unsampled location increases. The adaptive bandwidth \(h(x, y)\) is given by \(h(x, y) = c R_k(x, y)^2\), where $c$ is a parameter to control the bandwidth, and $R_k(x, y)$ is the average distance between $(x, y)$ and its $k$ nearest neighbors. Parameters $c$ and $k$ are both tuned using cross-validation.

\subsection{Evaluation}

\paragraph{Objective.} To assess whether the chosen methods can potentially yield accurate estimates at synthetic, out-of-sample grid locations, we first performed an in-sample evaluation. We thus evaluated the models on the preprocessed dataset using an 80-20 split for each month. Then, we compared the \textit{best case} latency estimate at each test location. We define the best case latency estimate as the one that minimizes the absolute error for a ground truth measurement conducted at that location. Finally, we compare the estimate and the ground truth, i.e, $\hat{Z}(x, y)$ and $ Z(x, y)$, across models. Our evaluation ensures that we use the same training and testing sets across all models and parameter choices.

\paragraph{Parameter Selection.} Our estimates were optimized using a grid search over each parameter choice. We set the parameters for each model as follows: $p = 1, 2, 3$ for IDW, and $span = 0.05, 0.1, 0.5, 1$ for LOESS. For STBKR, we vary $c$ between $10^{-5}$ and $100$, while $k$ is varied between $5$ and $1000$, both on logarithmic scales.

\begin{figure}[t]
  \centering
  \begin{subfigure}[b]{0.31\textwidth}
      \includegraphics[width=\textwidth]{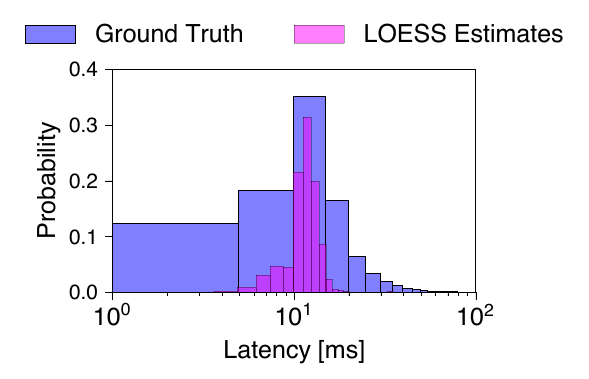}
      \caption{LOESS}
      \label{fig:error-loess}
  \end{subfigure}
  \hfill
  \begin{subfigure}[b]{0.31\textwidth}
      \includegraphics[width=\textwidth]{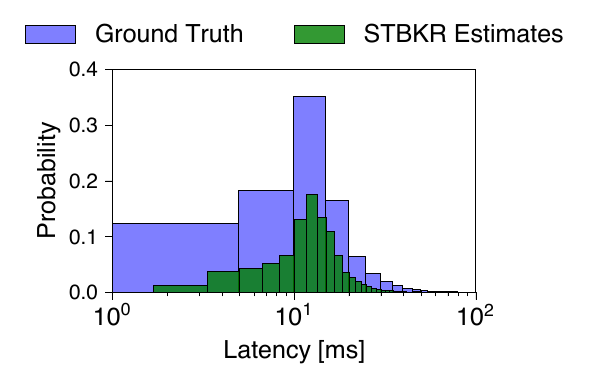}
      \caption{STBKR}
      \label{fig:error-stbkr}
  \end{subfigure}
  \hfill
  \begin{subfigure}[b]{0.29\textwidth}
      \includegraphics[width=\textwidth]{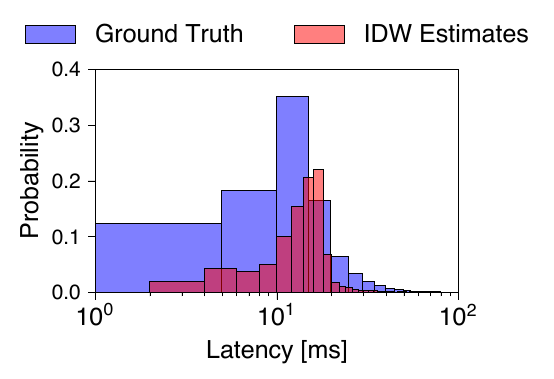}
      \caption{IDW}
      \label{fig:error-idw}
  \end{subfigure}
  \caption{Error analysis for prior interpolation techniques. The x-axis shows per-location latency values on a log scale. While STBKR provides well-aligned estimates, IDW shows a greater sensitivity to outliers in latency.}
  \label{fig:interpolation-errors}
\end{figure}

\paragraph{Overall Comparisons.} Figure~\ref{fig:interpolation-errors} shows the distributions of ground truth and predicted latency values across the three interpolation models. In Figure~\ref{fig:error-loess}, we observe a narrow range for LOESS estimates, suggesting that the model underestimates latency for a vast majority of locations. These underestimates are likely a result of the model's dependence on the locations' coordinates. LOESS performs a regression over the lat-long values directly. While this allows the model to capture the broader trends in latency across the geography, it is less effective in capturing the extreme values. So, using LOESS with a linear polynomial may not be appropriate when there is a greater presence of outliers in ground-truth latency estimates. In contrast, the STBKR model (Figure~\ref{fig:error-stbkr}) shows a slightly better alignment with the ground truth, suggesting that the model performs better than LOESS in capturing extreme values. However, in comparison to IDW, we notice that the distribution of STBKR estimates possesses a shorter rightmost tail, indicating that STBKR underestimates latency at locations with high ground-truth values. This is further confirmed by counting the number of locations for which we observe $> 50$ ms latency estimates for STBKR. We find that for IDW, the number of locations with $> 50$ ms latency is 2.04 times higher than STBKR. The ability to capture extreme values is crucial towards understanding the distribution of latency over a geography, especially when the focus is on identifying areas with poor connectivity. Further, STBKR being a Kernel regression method, has an $O(N^2)$ complexity for parameter tuning \cite{hastie2009elements}, where $N$ is the total number of data points. This makes it computationally expensive for large datasets. In contrast, IDW does not require an additional parameter tuning step, and involves lower number of computations, making it a suitable choice for large datasets such as Ookla.

\begin{takeawaybox}
  \paragraph{Takeaways.} We evaluate three deterministic interpolation techniques -- IDW, LOESS and STBKR -- to estimate latency by down-sampling the Ookla dataset. The lower computational complexity for IDW, coupled with its higher sensitivity to extreme values, makes it a suitable choice for large datasets such as Ookla. We thus choose IDW as the primary interpolation technique for our subsequent analysis.
\end{takeawaybox}
\section{Regionalization} \label{sec:regionalization}

% \begin{takeawaybox}
%     \textbf{Summary of findings:}
%     \vspace{1em}
%     \begin{itemize}[leftmargin=\parindent,align=left,labelwidth=\parindent,labelsep=0pt,topsep=0pt,itemsep=0pt]
%         \item We find the optimal number of clusters to be $N = 4$ for the hexagon cell overlay with a LOESS $span$ of 0.075 and the minimum number of clustering units kept to 50. The within and between-cluster variance for this choice of parameters is found to plateau for $N > 4$.
%         \item We observe significant visual differences between the clusters obtained for different sampling approaches, namely, raw neighborhoods, interpolated neighborhoods, and interpolated hexagon cells.
%         \item We evaluate the stability of our clusters by measuring the Jaccard similarity between clusterings for multiple smaller samples of the dataset. The median pairwise Jaccard similarity between these clusters is found to be 0.96 when we overlaid the city's map with hexagons. For neighborhood overlay, we find a similarity of 0.99 when we use prior interpolation, and 0.43 when we use raw measurement aggregates.
%     \end{itemize}
% \end{takeawaybox}

To construct the sampling boundaries for \chicago{}, we first interpolate latency measurements across the citywide geography at regular grid spacing. Then, we overlay these measurements with hexagon cells. Finally, we apply SKATER with aggregates calculated within these boundary units as the clustering metric. We next formally describe the problem and our approach to regionalization.

% \begin{figure*}[t]
%     \centering
%     \begin{subfigure}[b]{0.3\linewidth}
%         \centering
%         \includegraphics[width=\textwidth]{figures/sec5/span_75_ratio.pdf}
%         \caption{Sensitivity towards $floor$ at $span = 0.075$. \annotate{We obtain the lowest $\rho$ for $floor = 50$.}}
%         \label{fig:span-0075}
%     \end{subfigure}
%     ~
%     \begin{subfigure}[b]{0.3\linewidth}
%         \centering
%         \includegraphics[width=\textwidth]{figures/sec5/wb_ratio_tradeoff.pdf}
%         \caption{Within and between-cluster variance for $span = 0.075, floor = 50$ \annotate{plateaus for $N > 4$.}}
%         \label{fig:span-01}
%     \end{subfigure}
%     \caption{Parameter sensitivity analysis for $span = 0.075$ and $N$ in the range $\lbrack 2, 12 \rbrack$. We observe $N = 4$ to be optimal for the choice of $span = 0.075$ and $floor = 50$.}
%     \label{fig:reg-param-sens}
% \end{figure*}

\subsection{Problem Formulation}

The problem of discovering statistical sampling boundaries for latency can be restructured as an unsupervised learning problem. Consider a geographical region $\Omega$ and a set of its partitions $\mathcal{H} = \lbrace H_i \rbrace_{i = 1}^{n}$. Further, consider a set of latency measurements conducted over the region as $\mathcal{X} = \lbrace x_i \rbrace_{i = 1}^{n}$. Our goal is to find a set of spatially contiguous clusters $\mathcal{C} = \lbrace C_i \rbrace_{i = 1}^{N}$ such that each cluster $C_i$ is a subset of $\mathcal{H}$, and the latency values within each cluster are drawn from a common distribution. To achieve this, we calculate the mean latency for each partition $H_i$ as $\mu_i = \frac{1}{|H_i|} \sum_{x \in H_i} x$ and assign a feature vector $\mathbf{v}_i = \lbrack \mu_i \rbrack$ to each partition. To calculate the dissimilarity among the partitions, we consider the use of Euclidean distance, $d(\mathbf{v}_i, \mathbf{v}_j) = \|\mathbf{v}_i - \mathbf{v}_j\|$. Using this dissimilarity function, we apply a spatial clustering algorithm to group the partitions into $N$ clusters. Finally, we define the existence of a sampling boundary ($B$) between two partitions of the region $\Omega$ as:

\[
    B(H_i, H_j) = \begin{cases}
        \text{True,} & \text{if } H_i \in C_k~\text{and}~H_j \in C_l,~\text{with}~k \neq l \\
        \text{False,} & \text{otherwise}
    \end{cases}     
\]

\subsection{Technique}

We choose SKATER \cite{assunccao2006efficient} as the default regionalization algorithm. SKATER provides a fast and efficient way to identify spatially contiguous clusters in a given region. Additionally, it offers a way to control for homogeneity of the clusters by setting thresholds. The SKATER algorithm involves three main steps. First, it constructs a graph where each node represents a spatial unit, e.g, a census tract boundary, hexagonal units, geographic coordinates of Internet users, or a neighborhood. The edges between the nodes denote spatial adjacency, i.e., two nodes are connected if they share a common boundary. In case of points, the edges are constructed using a distance threshold. The weights of these edges are determined using the dissimilarity between the nodes, which is Euclidean distance in our case. In the second step, SKATER constructs a Minimum Spanning Tree (MST) from the graph. An MST is a tree that connects all the nodes in the graph using the minimum possible edge weights. The use of MST in this step ensures a faster runtime, as considering all edges in the graph is infeasible. In the final step, the MST is iteratively pruned by removing edges with the highest weights. This results in a set of connected components, one for each spatially contiguous cluster. The number and size of the clusters can be controlled using two parameters, $N$ and $floor$. $N$ denotes the number of clusters, and $floor$ denotes the minimum number of nodes in each cluster. We apply SKATER to the resolution-8 hexagon cell overlay for Chicago upon interpolating the latency values using IDW at regular grid spacing of 50 meters. 

\subsection{Cluster Optimization} \label{sec:cluster-optimization}

\paragraph{Objective.} Since there is little prior knowledge about the true number of clusters and their individual sizes, we perform a sensitivity analysis for optimizing the parameters for SKATER. There can be numerous ways to conduct this analysis. Akin to traditional clustering methods, approaches such as Silhouette score \cite{shahapure2020cluster} or Davies-Bouldin index \cite{karkkainen2000minimization} can be used to determine the optimal parameters for SKATER. While these approaches are widely used, they may give us a different set of parameters for each monthly set of IDW interpolated latency estimates. This is undesirable because true infrastructure boundaries may be less prone to changes over time. To address this issue, we use a more intuitive, grid-search based approach to find the optimal set of parameters. For each month, while varying $N$ and $floor$, we select the set of parameters that yield most similar boundaries between monthly fits. Further, for a given choice of $floor$, if increasing the number of clusters by one results in the same cluster boundaries, we cease increasing the number of clusters any further. This is because the additional cluster does not provide any additional information about the underlying spatial distribution of latency.

\begin{figure}[h]
    \centering
    \begin{subfigure}{0.4\linewidth}
        \includegraphics[width=\textwidth]{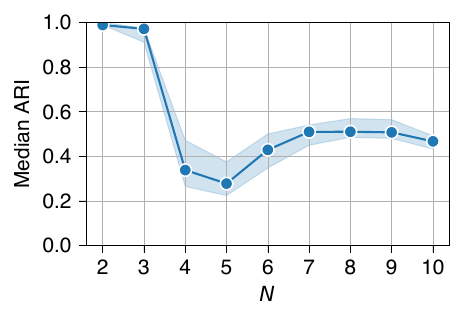}
        \caption{Median Adjusted Rand Index (ARI) as a function $N$.}
        \label{fig:mean-ari-floor}
    \end{subfigure}
    \hfill
    \begin{subfigure}{0.3\textwidth}
        \includegraphics[width=\linewidth]{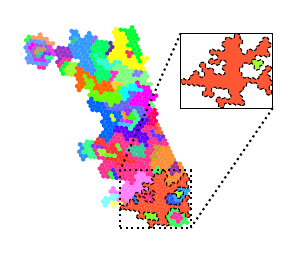}
        \caption{Boundaries for May 2022 with $N = 77$. Zoomed-in green region is a spatial outlier.}
        \label{fig:example-clusters}
    \end{subfigure}
    \hfill
    \begin{subfigure}{0.22\textwidth}
        \includegraphics[width=\linewidth]{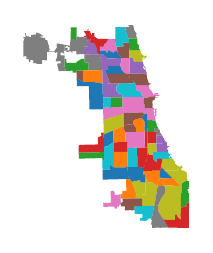}
        \caption{Boundaries for 77 neighborhoods in Chicago.}
        \label{fig:neighborhood-boundaries}
    \end{subfigure}
    \caption{Analysis of clustering performance using SKATER. \ref{fig:mean-ari-floor} shows the median ARI score for $floor = 2$ against $N$ calculated over monthly fits of SKATER. \ref{fig:example-clusters} and \ref{fig:neighborhood-boundaries} compare the resulting clusters for $N = 77$ and $floor = 2$ with the neighborhood boundaries for Chicago. Boundaries drawn from measurement data do not align with administrative boundaries.}
\end{figure}

\paragraph{Parameter Selection.} We use the Adjusted Rand Index (ARI) to compare the clusterings obtained for different choices of $N$ and $floor$. For each parameter combination, we calculate the median ARI score between each pair of monthly clusterings. We generally observed a low sensitivity of the ARI towards the choice of $floor$ (varied between 2 and 37) when we used mean to aggregate latency over each cell. Figure~\ref{fig:mean-ari-floor} shows the median ARI score for $floor = 2$ versus $N$ calculated over monthly fits of SKATER. We choose a $floor$ of 2 because it helps us identify small clusters with significant differences in latency from their neighbors. We observe that a lower $N$ ($N \leq 3$) allows for more stable clusters, but the similarity between the resulting clusters tends to stabilize between $N = 7$ and $N = 9$. Beyond this stage, the median ARI starts to decline as the clusters become increasingly fragmented. We choose $N = 7$ and $floor = 2$ for subsequent analysis and demonstrations as this combination gives a higher similarity in the vicinity of the stabilization point. This choice also allows us to identify fine-grained boundaries while ensuring that the clusters are not overly fragmented.

\paragraph{Misalignment with Administrative Boundaries.} We also check whether data-driven boundaries generated using SKATER align with administrative area boundaries. In Figure~\ref{fig:example-clusters} and Figure~\ref{fig:neighborhood-boundaries}, we compare the resulting clusters for $N = 77$ and $floor = 2$ with the neighborhood boundaries for Chicago. The choice of $N$ is the same as the total number of neighborhoods in Chicago. We observe that the clusters drawn from latency measurement data do not show a one-to-one correspondence with the administrative boundaries. This result suggests that sampling along administrative boundaries may not be the best approach for understanding the spatial distribution of latency.

\begin{takeawaybox}
\paragraph{Takeaways.} While we choose $floor = 2$ and $N = 7$ for subsequent analysis, we argue that the choice of $N$ and $floor$ ultimately depends on the policy use-case under consideration. For a given $floor$, a lower value of $N$ may be suited in scenarios that may involve allocating a limited investment budget towards larger divisions of the geography. A higher value of $N$, instead, is desirable in cases when targeted interventions may be required in regions marked by abnormal performance. An example of such a region is the green region shown in the zoomed-in box in Figure~\ref{fig:example-clusters}. In comparison with the citywide average of 19.07 ms, we observe a higher average latency of 45.04 ms for this region. The orange neighboring region shows an average latency of 17.15 ms. This also demonstrates the ability of our approach in identifying spatial outliers in latency. 
\end{takeawaybox}

\section{Cluster Sensitivity to Data Sampling Choices} \label{sec:sample-effects}

Our analysis thus far used a single, IDW-interpolated sample of Ookla measurements in combination with average latency as the default clustering metric. In this section, we discuss the sensitivity of our results towards the choice of sampling and aggregation methods. We experiment with the aggregation metric, the aggregation unit, and the ISP to compare sampling boundaries. 

\subsection{Impact of Aggregation Metric} \label{sec:agg-metric}

\begin{figure}[ht]
    \centering
    \begin{subfigure}[b]{0.48\textwidth}
        \includegraphics[width=\textwidth]{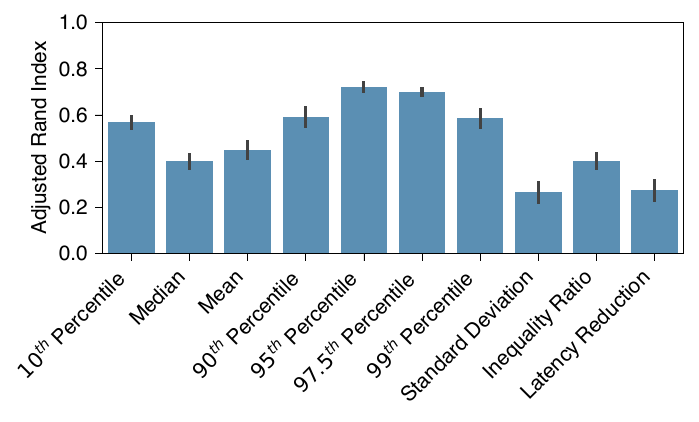}
        \caption{$floor = 2$, $N = 7$}
        \label{fig:floor_2}
    \end{subfigure}
    \begin{subfigure}[b]{0.48\textwidth}
        \includegraphics[width=\textwidth]{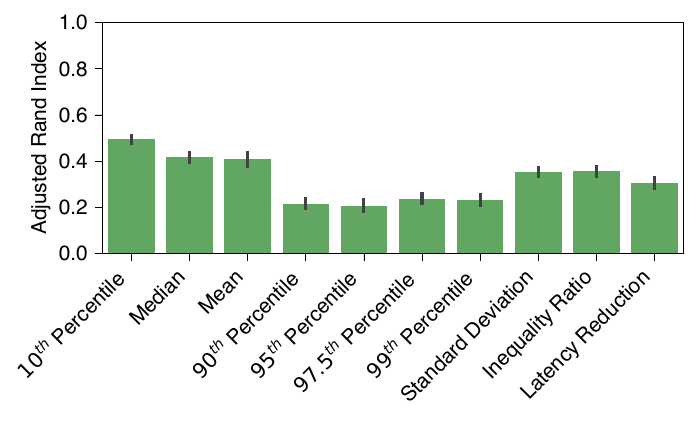}
        \caption{$floor = 25$, $N = 7$}
        \label{fig:floor_25}
    \end{subfigure}
    \caption{Comparison of boundary similarities under two extreme values of $floor$. Higher percentiles show a greater ARI score when we require more homogeneous clusters.}
    \label{fig:impact-of-metric}
\end{figure}

\paragraph{Experiment Setup.} We first assess \textit{which aggregation metrics lead to more stable boundaries across monthly fits}. While mean, percentiles and standard deviation are more intuitive metrics to pick from, a number of additional, compound metrics have been proposed in prior work. The \textit{inequality ratio} \cite{martin2023divided} is defined as the ratio between the $90^{th}$ and $10^{th}$ percentile of latency. A higher value of this metric indicates a higher degree of variability in the latency measurements within a hexagonal cell. Latency reduction \cite{sharma2022benchmarks}, i.e, the difference between $90^{th}$ and $10^{th}$ percentile latency, is also considered for stability checks. For each choice of metric, we calculate the ARI score between the boundaries obtained using SKATER. Then, we compare the distribution of pairwise ARI scores calculated between monthly fits of the algorithm.

\paragraph{Observations.} Figure~\ref{fig:impact-of-metric} shows a comparison of the pairwise ARI scores for above metric choices. In Figure~\ref{fig:floor_2}, with a $floor$ of 2, we observe that the $10^{th}$ and the $90^{th}$ percentile show a similar ARI score, while the ARI score for the $95^{th}$ and $97.5^{th}$ percentiles is higher. Contrary to this, in Figure~\ref{fig:floor_25}, we observe a reduced stability as we move from the $10^{th}$ to the $90^{th}$ percentile while using a higher value of $floor$. Other metrics show moderate levels of ARI scores. 

\paragraph{Dependence on cluster size.} We argue that above observations are an artifact of an important trade-off. The $floor$ parameter, in essence, controls the balance between homogeneity and contiguity in the resulting clusters. A lower $floor$ results in clusters of greater homogeneity but less contiguity, while a higher $floor$ results in larger but less homogeneous clusters. When $floor$ is set to 2, a superior homogeneity in the clusters reduces the difference between the $10^{th}$ and $90^{th}$ percentiles, leading to similar ARI scores for these metric choices in a cluster. Since the cluster sizes are relatively small, regions showing consistently high levels of latency are likely to persist across months. This leads to a higher similarity for the $95^{th}$ and $97.5^{th}$ percentiles. The $99^{th}$ percentile is more likely to capture rare events such as outages or congestion, and thus registers a lower ARI score than the $97.5^{th}$ percentile. When $floor$ is changed to 25, the clusters become more contiguous and large, with greater differences arising between the extreme values. Across larger areas, the baseline latency is likely to remain stable, whereas the variability in latency above the baseline is expected to be greater. This leads to a reduced similarity as we move from the $10^{th}$ to the $90^{th}$ percentile.

\begin{takeawaybox}
    \paragraph{Takeaways.} The above results are relevant from a policy standpoint because they highlight the importance of choosing the right aggregation metric for carving out sampling boundaries for latency. When the policy objectives are to identify smaller regions with consistently poor latency for targeted interventions, a lower $floor$ coupled with a higher percentile metric such as the $90^{th}$ or $95^{th}$ is more suitable. On the other hand, when there is a need to identify larger regions with varying levels of latency, a higher $floor$ in combination with a lower percentile metric, such as the $10^{th}$, is more preferable.
\end{takeawaybox}

\subsection{Impact of Aggregation Unit}\label{sec:agg-unit}

\begin{figure}[ht]
    \centering
    \begin{subfigure}[b]{0.3\textwidth}
        \includegraphics[width=\textwidth]{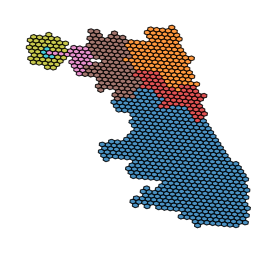}
        \caption{Regular Hexagons}
        \label{fig:hex_bdry}
    \end{subfigure}
    \begin{subfigure}[b]{0.3\textwidth}
        \includegraphics[width=\textwidth]{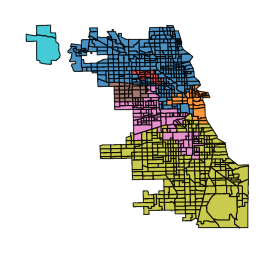}
        \caption{Census Tracts}
        \label{fig:ct_bdry}
    \end{subfigure}
    \begin{subfigure}[b]{0.3\textwidth}
        \includegraphics[width=\textwidth]{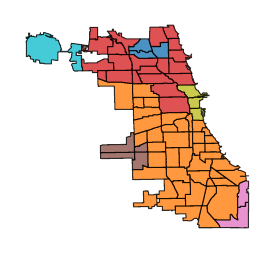}
        \caption{Neighborhoods}
        \label{fig:nbh_bdry}
    \end{subfigure}
    \caption{Comparison of boundaries under different aggregation unit choices with $N = 7$ for June 2022. The
    choice of sampling unit can significantly affect resulting sampling boundaries, and hence our conclusions about the spatial distribution of latency.}
    \label{fig:boundary_comparison}
\end{figure}

\paragraph{Experiment Setup.} We next evaluate how the choice of the smallest spatial unit affects the similarity among monthly fits of SKATER. To this end, we consider the use of three unit types -- regular hexagons, census tracts and neighborhood boundaries for Chicago. We continue with our choice of $N = 7$ but set $floor$ to 1 for these comparisons. This is because for census tracts, we noticed that two of the tracts were disconnected from the rest, leading to two distinct spatial islands. Due to this behavior, we notice that any choice with $floor \geq 2$ led to the exclusion of these disconnected tracts from the analysis, potentially skewing the results and reducing the validity of our comparisons. By setting $floor$ to 1, we ensure that all spatial units, however small, are included in the analysis, and we make fair comparisons between the three unit choices.

\begin{figure}
    \centering
    \begin{subfigure}{0.45\textwidth}
        \includegraphics[width=\textwidth]{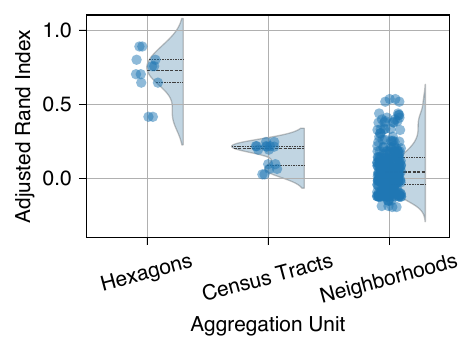}
        \caption{Raw Averaging}
        \label{fig:unit-ari-raw}
    \end{subfigure}
    \hfill
    \begin{subfigure}{0.45\textwidth}
        \includegraphics[width=\textwidth]{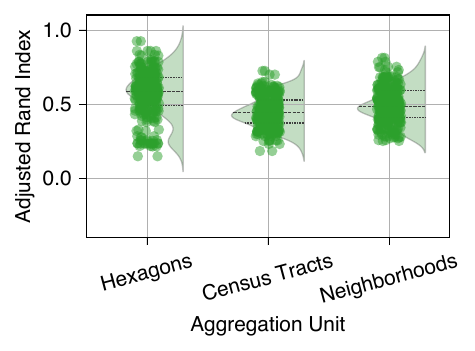}
        \caption{Interpolated Averaging}
        \label{fig:unit-ari-interpolated}
    \end{subfigure}
    \caption{Pairwise ARI comparison across aggregation unit choices and averaging method for raw and interpolated data. Raw averaging is only suitable when computed over regular and small spatial scales.}
    \label{fig:unit-ari-comparison}
\end{figure}

\paragraph{Observations.} Figure~\ref{fig:boundary_comparison} shows an example comparison between the resulting boundaries for June 2022. We notice that the South and South-West sides are clustered together in all three cases, with this boundary extending up to the far North for neighborhood units. Additionally, we observe a greater degree of variability in the Northern regions, with the hexagonal units providing more fine-grained sets of clusters. To further understand the cluster quality for the above unit choices with or without interpolation, we plot the monthly pairwise ARI score distributions in Figure~\ref{fig:unit-ari-raw} and Figure~\ref{fig:unit-ari-interpolated}. We see that prior interpolation ensures a consistent ARI score across the choice of aggregation units, with hexagons achieving the highest median ARI scores. Additionally, we see that raw averaging of point latency measurements tends to produce worse clustering performance in the case of census tract and neighborhood units. Overall, hexagonal units with post-interpolation averaging register a median ARI score of 0.59, in comparison to a score of 0.20 for the traditional practice of using census tract units with raw averaging.

\paragraph{Unstable administrative clusters with raw averaging.} Census tracts and neighborhood boundaries tend to be unevenly distributed in size, shape and population density, leading to irregular smoothing of latency variations, and thus less stable boundaries. For hexagonal cells, we observe a superior clustering performance for raw averaging, suggesting that raw averaging is only reliable when computed over small yet regular spatial units. Interpolated averaging, on the other hand, ensures consistent boundaries across the choice of aggregation units, with hexagons showing the highest median ARI scores. This is because prior interpolation helps capture the underlying spatial trends in the data, leading to more consistent boundaries across different spatial unit choices.

\begin{takeawaybox}
    \paragraph{Takeaways.} Above results signal a significant departure from the status quo of using raw averaging for aggregating Internet performance over administrative regions. For policy, it is thus advisable to use prior interpolation to ensure consistency in results when using administrative boundaries, or high resolution, regular spatial units if computing aggregates directly from raw measurements.
\end{takeawaybox}

\subsection{Impact of ISP}\label{sec:isp-effects}

\begin{figure}
    \includegraphics[width=0.45\textwidth]{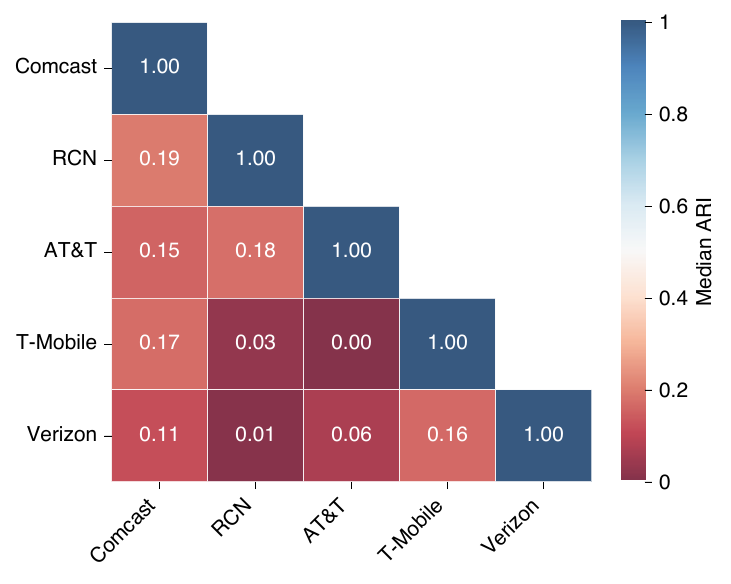}
    \caption{Heatmap of ARI scores between ISPs for the 17-month duration. We observe low similarity between the boundaries obtained for different ISPs.}
    \label{fig:isp-ari}
\end{figure}

\paragraph{Experiment Setup.} Our samples thus far have included latency data from multiple ISPs together for aggregation into cells. We now evaluate \textit{how much do ISPs share latency boundaries among their respective samples.} We therefore first select top-five ISPs on the basis of their measurement counts. Then, we apply IDW individually to the 17-month-long samples on a per-ISP, per-month basis to interpolate these measurements over the city. Next, with a $floor = 2$ and $N = 7$, we apply SKATER after aggregating these measurements over hexagonal cells using the $95^{th}$ percentile. This choice was made to ensure that we compare the most homogeneous clusters between ISPs using our best case metric from Section~\ref{sec:agg-metric}. Finally, we calculate the median ARI score between the boundaries obtained for each ISP pair over the 17-month period.

\paragraph{Observations.} Figure~\ref{fig:isp-ari} shows a heatmap of the resulting ARI scores between the top five ISPs. We observe that the ISPs generally show a lower degree of similarity between their sampling boundaries. In contrast, when we interpolate measurements over a collective sample derived from all ISPs (as shown previously), we observe a higher degree of similarity between the boundaries. 

\paragraph{Low similarity between ISP boundaries.} When using a collective sample, the aggregation process smoothens out variability in latency measurements across ISPs, leading to more consistent boundaries over time. The differences between ISPs may arise from several factors such as the geographic distribution of servers, the underlying access technology, the network design for the ISPs, or the per-ISP sample size. When measurements from all ISPs are jointly interpolated, individual ISP characteristics are more likely to be smoothened out, and the resulting boundaries are more likely to reflect the geographic distribution of latency in the region. \rev{Therefore, a one-size-fits-all approach to sampling boundaries may not be suitable for all ISPs with the current method.}{Looking at each ISP individually on a regionalized map is likely to provide a more consumer-transparent view of network performance in a region. It would not only help new subscribers make informed decisions about their Internet connections, but also assist in maintaining greater scrutiny over ISPs regarding their service level agreements (SLAs). The FCC should thus consider releasing ISP and region specific maps to ensure that the representations of network performance are accurate and reliable.}

\begin{takeawaybox}
    \paragraph{Takeaways.} If a network operator is interested in understanding the spatial distribution of latency for their network, they should not rely on boundaries constructed from a heterogeneous sample from multiple ISPs. Rather, they are recommended to use their own data to make informed decisions about infrastructure upgrades. Moreover, funding agencies and regulators should consider the ISP-specific nature of latency when making decisions about infrastructure investments.
\end{takeawaybox}
\section{Improving Clustering Stability} \label{sec:volatility}

 While our use of ARI score as an evaluation metric allows for understanding the degree of similarity between boundaries, it does not help localize the boundary variation, or volatility in each clustering. Localizing this volatility can be crucial for making informed decisions about infrastructure investments. Areas with high clustering volatility can be used for running additional measurement campaigns to fill data gaps, while those showing low volatility can be prioritized for immediate policy interventions. We use bootstrap resampling \cite{efron1994introduction} techniques to calculate this metric. Bootstrap resampling is often used to estimate sample statistics when the underlying distribution of the data is unknown. In our case, we use this technique to simulate the perturbations across clusters to estimate the distribution of volatility in cluster assignments. We define volatility for a cell as the probability that applying SKATER on two different samples of the same dataset will result in different cluster assignments for the cell. Formally, for a cell $H_i$, we define the clustering volatility as:

    \[
        V(H_i) = \frac{\sum_{j = 1}^{|B|} \sum_{k = 1}^{|B|} I\{\mathcal{C}_{j}(H_i) \neq \mathcal{C}_{k}(H_i)\}}{\binom{|B|}{2}}
    \]

where $|B|$ denotes the number of bootstrap samples drawn from the dataset. A total of 1000 samples is often considered adequate for most practical use cases \cite{efron1994introduction}, so we choose $|B| = 1000$. $\mathcal{C}_j (H_i)$ denotes the cluster assignment for the cell $H_i$ when SKATER with $N = 7$ and $floor = 2$ is applied to the $j^{th}$ bootstrap sample. $I\{\mathcal{C}_{j}(H_i) \neq \mathcal{C}_{k}(H_i)\}$ denotes the indicator function that assumes 1 when two different bootstrap samples result in a different cluster assignment, and 0 otherwise. $\binom{|B|}{2}$ denotes the binomial coefficient, indicating the total number of pairs of cluster assignments for the cell.

Bootstrap resampling in its original form assumes that the samples drawn are independently and identically distributed. For calculating clustering volatility, we use a spatial version of this technique, called block bootstrapping \cite{radovanov2014comparison}. Block bootstrapping accounts for potential spatial auto-correlation in the data that may be induced due to smoothing of variations caused by interpolation. We estimate that the interpolated dataset produces a global Moran's I \cite{moran1950notes} (an indicator of overall spatial dependence) of 0.795, on a scale of -1 and 1, with a significance level of 0.001 among $10^{th}$ percentile latency aggregates, suggesting that the smoothing caused by IDW induced significant spatial auto-correlation. In block bootstrapping, instead of drawing individual point samples independently, we thus resample points from complete hexagonal cells to account for local spatial dependencies in the data. We use a block size of one cell under the assumption that spatial dependencies between cells are minimal. We then calculate the clustering volatility for each cell over the city.

\begin{figure}
    \includegraphics[width=0.75\textwidth]{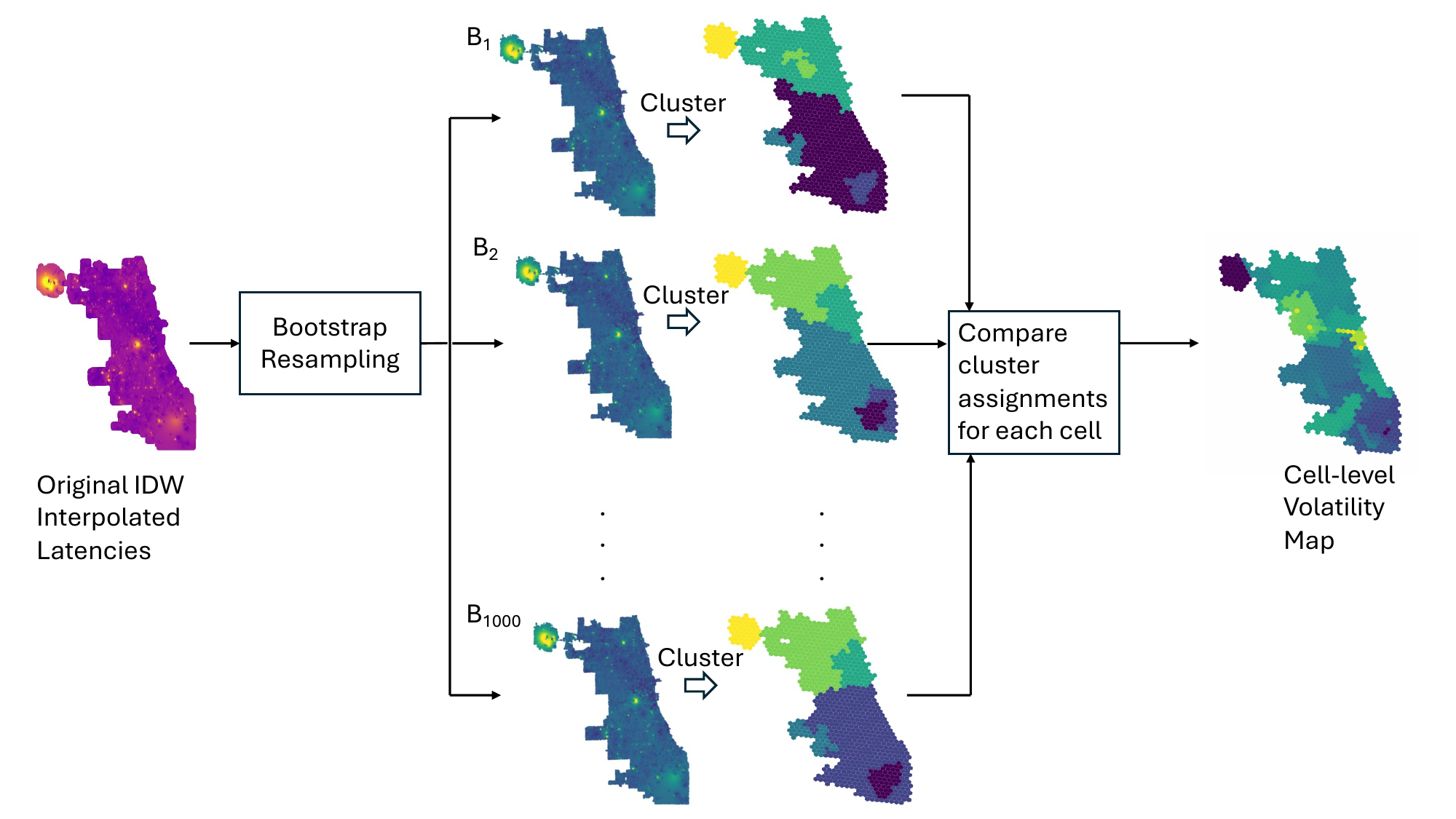}
    \caption{Procedure for computing a map of clustering volatility for Chicago.}
    \label{fig:boundary-alignment}
\end{figure}

\begin{figure}
    \includegraphics[width=0.3\textwidth]{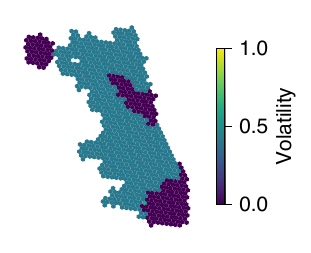}
    \caption{Resulting clustering volatility map for January 2022. The $10^{th}$ percentile of latency is used as the aggregation metric. Block bootstrapping identifies three distinct regions of zero boundary volatility.}
    \label{fig:volatility-block-bootstrap}
\end{figure}

Figure~\ref{fig:volatility-block-bootstrap} shows an example boundary volatility map for Chicago for January 2022 with $10^{th}$ percentile as the aggregation metric under block bootstrapping. We use $N = 7$ and $floor = 2$ for this analysis. We observe that majority of the hexagonal cells show a low-to-moderate level of volatility. These regions show some fuzziness in the boundaries, indicating that the cluster assignments for these cells are more likely to change across different samples of the interpolated dataset. We also observe that three distinct regions show zero boundary volatility overall. The North-Western region hosts the city's airport, the Central-Eastern region is marked by Chicago's central business district (CBD), while the Southern region generally represents areas with a high poverty rate \cite{kaufman2013chicago}. These areas are likely to represent distinct network conditions than other suburban areas, leading to more stable boundaries over time. Regions falling in these contiguous cells with low volatility can be prioritized for immediate policy interventions, depending on the underlying latency distribution. On the other hand, regions marked with significant volatility, such as some suburban regions, can benefit from additional data collection efforts for ensuring more stable boundaries.

\section{Related Work}\label{sec:related}

% Aspatial studies -- no spatial information

Crowdsourced speed test datasets, such as Ookla and M-Lab have found a variety of policy use cases in prior research. Bauer \etal~\cite{bauer2010understanding} describe the best practices for reporting data to reason about advertised and measured speeds for ISPs. Similarly, Feamster and Livingood \cite{feamster2020measuring} describe the need to augment speed test outputs with additional contextual information to increase the scope and usability of crowdsourced data. Going a step further, Paul \etal~\cite{paul2022importance} showed an empirical characterization of several factors that may confound Internet performance of an access link. Further, Macmillan \etal~\cite{macmillan2023comparative} augment real-world speed test data with laboratory experiments to understand the differences between M-Lab's NDT and Ookla's Speedtest tools. They find that Ookla Speedtest tends to report higher speeds than M-Lab NDT under high latency conditions, further showcasing the need for additional context in analyzing speed test data. Finally, Clark and Wedeman \cite{clark2021measurement} discuss the need to interpret aggregate statistics on speed test data over geographies to understand the overall behavior of the Internet. While these studies collectively point towards a need for a comprehensive understanding of the Internet access landscape, they do not leverage the spatial nature of the crowdsourced measurement data to inform policy decisions. 

% Hybrid studies -- those that leverage some spatial information

More recently, there is an emergence of studies that do leverage location information from crowdsourced speed test data for analysis. Paul \etal~\cite{paul2022characterizing} summarize upload and download speeds within census block groups using the M-Lab dataset, and correlate these speeds with demographic data. They find income levels to possess significant relationship with download speed. Further, Lee \etal~\cite{lee2023analyzing} present a methodology for regional bias correction in crowdsourced Ookla speed test measurements. Krzysztof \etal~\cite{janc2022spatial} explore local Internet quality in Poland using a spatial analysis on an Ookla dataset, highlighting the non-uniformity in access quality among rural and urban areas. Caldas \etal~\cite{caldas2023assessing} perform a similar analysis over Denmark to highlight Internet access disparities in Denmark. Though these studies leverage spatial information, they: (1) still rely on aggregates calculated over predefined social boundaries for analysis, and (2) do not take into account uneven densities of measurements within these boundaries. These limitations can lead to coarse generalizations, and can prevent policymakers from identifying specific areas that may need immediate interventions.
 
% Studies that use some interpolation techniques

Another class of studies closely related to our work directly leverage point measurements by not assuming prior structure to measurement sampling boundaries. For instance, Sommers \etal~\cite{sommers2012cell} use Inverse Distance Weighting (IDW) to understand the spatial distribution of cellular and Wi-Fi performance in metro areas. Similar to Caldas \etal, they observe a degradation in performance as one moves further away from metro areas. Jiang \etal~\cite{jiang2023mobile} propose the Self-tuning Bandwidth in Kernel Regression (STBKR) technique to estimate cellular speed test quality using speed measurements from Ookla. They find that STBKR outperforms Kriging in accurately estimating throughput in sparsely sampled regions. The LOESS technique, used in astronomy \cite{Cappellari2013b} to analyze the trajectories of celestial objects, has not been used in the context of Internet performance. In our work, we extend these techniques to identify distinct regions for sampling Internet performance on the basis of latency, with a focus on stability of these boundaries over time and sampling variations.

% how does our work differ from prior work

% We find very few studies with a notable use of spatial interpolation in prior Internet measurement literature. The LOESS technique has been used in some spatial applications in astronomy \cite{Cappellari2013b}, but not in the context of Internet performance. IDW found its use in Sommers \etal~\cite{sommers2012cell} to understand the spatial distribution of cellular and Wi-Fi performance in metro areas. They observed a degradation in performance as one moves further away from metro areas. Stuyvesant \etal~\cite{stuyvesant2023michigan} used Kriging to estimate the spatial distribution of broadband speeds in Michigan. Jiang \etal~\cite{jiang2023mobile} proposed the STBKR technique but applied it on cellular speed test data. Our work differs from these studies in that it focuses primarily on latency measurements, and presents a comparison between deterministic interpolation methods.

\label{lastpage} \section{Conclusion \& Future Work}\label{sec:conclusion}

This work presents a new approach for discovering statistical latency sampling
boundaries within a geographic region, such as a city, using crowdsourced
latency measurements. The findings of this study underscore the importance of
spatial analysis in network planning and the benefits of targeted
infrastructure investments for equitable Internet access. We show that the
method we develop can identify contiguous geographic regions with poor
Internet performance; such information can be used to inform policy
interventions and also assist ISPs with infrastructure planning. We summarize
the implications of our work and discuss potential future directions below.

\paragraph{Applying our approach to policymaking.} Our work provides a method
for identifying Internet latency sampling boundaries, assuming minimal
information about the underlying infrastructure. By delineating clear latency
boundaries, network operators can target specific areas for infrastructure
improvements, optimizing resource allocation and enhancing overall network
performance. \rev{Policymakers can apply these insights to develop regulations
and incentives that promote equitable access to low latency internet across
different regions, and addressing the digital divide.}{For instance, our illustration from Figure~\ref{fig:example-clusters} provides a starting point for identifying areas with higher latency. Given a sample of diagnostic measurement data and a specific timescale, contiguous regions with poor latency can be identified over a geography in a similar manner. Even though network operators may have their own measurement infrastructure in place, our method is particularly valuable for leveraging crowdsourced data, which can provide coverage from a large number of real-world vantage points in a cost-effective manner. It enables operators to work with scattered samples, where continuous measurements are unavailable. Integrating our clustering approach with further diagnostic information can give rise to suitable infrastructure improvements such as cable upgrades, deployment of additional hardware, or routing optimizations in under-provisioned areas. Our finding from Section~\ref{sec:agg-unit} suggests that the use of regular spatial units in place of administrative boundaries is expected to generate Internet performance representations that are more appropriate to inform long-term policy interventions. Finally, our finding from Section~\ref{sec:isp-effects} provides an important insight into using crowdsourced data for the FCC. Instead of relying on a large heterogeneous dataset, regulators can use our method to identify boundaries on a per-ISP basis, which can help in understanding the impact of different ISPs on overall network performance in a region.} 

\rev{}{\paragraph{Reliance on crowdsourced measurement data.}} Although our approach helps
reveal significant spatial structure from latency measurements, our sampling
boundaries are still based on crowdsourced data. Despite \rev{of}{} its wide adoption,
Ookla data may not be representative of all users, especially those with
limited or no Internet access. Future work can thus explore the possibility
of exploring an augmentation of multiple data sources using novel metrics to
improve the representativeness of the data. Our findings from
Section~\ref{sec:agg-metric} provide a starting point for such an exploration.
The design of spatial clustering methods that \rev{alleviate}{improve} the ARI scores for
metrics such as the inequality ratio, or latency reduction, can be a promising
direction, \rev{}{as these metrics can potentially remain consistent across datasets and device types.} \rev{Validating these boundaries using ground truth data from network
operators and regulators to ensure their accuracy and reliability is another
important future direction.}{}

\paragraph{Sensitivity to interpolation.} A key component of our method relies
on prior interpolation methods to build a surface model of latency across the
city. Although we primarily use IDW interpolation in this work, \rev{it remains to
be seen how much boundaries produced by different interpolation methods agree}{the mutual agreement between boundaries drawn using different interpolation algorithms is remaining to be evaluated}.
Additionally, interpolation methods that incorporate local context such as
network topology, routing information, and urban infrastructure can be
explored to improve the accuracy of the surface model. This 
information can enhance the stability of the clustering boundaries and
the accuracy of the regionalization process.

\paragraph{Geographic scope.} The primary focus of our analysis is on the city
of Chicago. \rev{}{Our approach identifies boundaries for sampling Internet latency without relying on Chicago-specific information such as census or demographic data.} While Chicago's diverse urban environment offers an intriguing
testbed, we expect our methodology to be generalizable to other cities with
similar data availability and population density. A key challenge to extending
our work to rural and remote areas with low connectivity is the availability
of crowdsourced data. Operators and regulators can play an important role in
collecting and sharing data from these regions to explore the potential of our
methodology in these areas.
 
\paragraph{Temporal analysis.} Our criterion for identifying sampling
boundaries from latency data is based on the stability of the sampling
boundaries across multiple temporal samples. While our distance calculations
for prior interpolation involve geographic distances only, our methods can be
extended to temporal distances as well. Such an approach would allow a more
nuanced understanding of the stability of boundaries, and the impact of
temporal variations in network performance on the sampling boundaries.

\section*{Ethics}

In this work, we analyze a proprietary Ookla dataset under a data usage agreement (DUA). In this dataset, the geolocations of Ookla users were truncated upto 4 decimal places, which allows a margin of a few hundred meters. The IP addresses are masked up to the last octet, which ensures anonymity. We did not find any other personally identifiable information in the dataset. Our research, therefore, does not raise any ethical concerns.

\section*{Acknowledgments}

We thank the anonymous reviewers and our shepherd Paul Barford for their constructive feedback and guidance. We are grateful to Jared Schachner, Jonatas Marques, and other colleagues at the Internet Equity Initiative (IEI) for their feedback on this research. This work was supported by National Science Foundation (NSF) grants IMR-2319603, CCRI-2213821, and SAI-2324515 at the University of Chicago, and by the ANR Project No ANR-21-CE94-0001-01 (MINT) at ENS de Lyon.

\bibliographystyle{ACM-Reference-Format}
\bibliography{paper}


\begin{thebibliography}{61}

%%% the Web address.

\ifx \showCODEN    \undefined \def \showCODEN     #1{\unskip}     \fi
\ifx \showDOI      \undefined \def \showDOI       #1{#1}\fi
\ifx \showISBNx    \undefined \def \showISBNx     #1{\unskip}     \fi
\ifx \showISBNxiii \undefined \def \showISBNxiii  #1{\unskip}     \fi
\ifx \showISSN     \undefined \def \showISSN      #1{\unskip}     \fi
\ifx \showLCCN     \undefined \def \showLCCN      #1{\unskip}     \fi
\ifx \shownote     \undefined \def \shownote      #1{#1}          \fi
\ifx \showarticletitle \undefined \def \showarticletitle #1{#1}   \fi
\ifx \showURL      \undefined \def \showURL       {\relax}        \fi
% invisible to TeX
\providecommand\bibfield[2]{#2}
\providecommand\bibinfo[2]{#2}
\providecommand\natexlab[1]{#1}
\providecommand\showeprint[2][]{arXiv:#2}

\bibitem[Aldworth and Cressie(1999)]%
        {aldworth1999sampling}
\bibfield{author}{\bibinfo{person}{Jeremy Aldworth} {and} \bibinfo{person}{Noel
  Cressie}.} \bibinfo{year}{1999}\natexlab{}.
\newblock \showarticletitle{Sampling designs and prediction methods for
  Gaussian spatial processes}.
\newblock In \bibinfo{booktitle}{\emph{Multivariate analysis, design of
  experiments, and survey sampling}}. \bibinfo{publisher}{CRC Press},
  \bibinfo{pages}{25--78}.
\newblock


\bibitem[Anselin(2018)]%
        {anselin2018spatial}
\bibfield{author}{\bibinfo{person}{Luc Anselin}.}
  \bibinfo{year}{2018}\natexlab{}.
\newblock \showarticletitle{Spatial Clustering (2)}.
\newblock \bibinfo{journal}{\emph{Disponible en}} (\bibinfo{year}{2018}).
\newblock


\bibitem[Assun{\c{c}}{\~a}o et~al\mbox{.}(2006)]%
        {assunccao2006efficient}
\bibfield{author}{\bibinfo{person}{Renato~M Assun{\c{c}}{\~a}o},
  \bibinfo{person}{Marcos~Corr{\^e}a Neves}, \bibinfo{person}{Gilberto
  C{\^a}mara}, {and} \bibinfo{person}{Corina da Costa~Freitas}.}
  \bibinfo{year}{2006}\natexlab{}.
\newblock \showarticletitle{Efficient regionalization techniques for
  socio-economic geographical units using minimum spanning trees}.
\newblock \bibinfo{journal}{\emph{International Journal of Geographical
  Information Science}} \bibinfo{volume}{20}, \bibinfo{number}{7}
  (\bibinfo{year}{2006}), \bibinfo{pages}{797--811}.
\newblock


\bibitem[{Battle For the Net}(2022)]%
        {BattleForTheNet2022}
\bibfield{author}{\bibinfo{person}{{Battle For the Net}}.}
  \bibinfo{year}{2022}\natexlab{}.
\newblock \bibinfo{booktitle}{\emph{{Internet Health Test based on Measurement
  Lab NDT}}}.
\newblock
\urldef\tempurl%
\url{https://www.battleforthenet.com/internethealthtest/}
\showURL{%
\tempurl}


\bibitem[Bauer et~al\mbox{.}(2010)]%
        {bauer2010understanding}
\bibfield{author}{\bibinfo{person}{Steven Bauer}, \bibinfo{person}{David~D
  Clark}, {and} \bibinfo{person}{William Lehr}.}
  \bibinfo{year}{2010}\natexlab{}.
\newblock \showarticletitle{Understanding broadband speed measurements}. Tprc.
\newblock


\bibitem[Bischof et~al\mbox{.}(2017)]%
        {bischof2017characterizing}
\bibfield{author}{\bibinfo{person}{Zachary~S Bischof},
  \bibinfo{person}{Fabian~E Bustamante}, {and} \bibinfo{person}{Nick
  Feamster}.} \bibinfo{year}{2017}\natexlab{}.
\newblock \showarticletitle{Characterizing and improving the reliability of
  broadband internet access}.
\newblock \bibinfo{journal}{\emph{arXiv preprint arXiv:1709.09349}}
  (\bibinfo{year}{2017}).
\newblock


\bibitem[{Broadband Internet Technical Advisory Group (BITAG)}(2022)]%
        {bitag_latency_explained}
\bibfield{author}{\bibinfo{person}{{Broadband Internet Technical Advisory Group
  (BITAG)}}.} \bibinfo{year}{2022}\natexlab{}.
\newblock \bibinfo{booktitle}{\emph{Latency Explained}}.
\newblock
\urldef\tempurl%
\url{https://www.bitag.org/documents/BITAG_latency_explained.pdf}
\showURL{%
\tempurl}


\bibitem[Caldas et~al\mbox{.}(2023)]%
        {caldas2023assessing}
\bibfield{author}{\bibinfo{person}{Maria~Paula Caldas}, \bibinfo{person}{Paolo
  Veneri}, {and} \bibinfo{person}{Michelle Marshalian}.}
  \bibinfo{year}{2023}\natexlab{}.
\newblock \showarticletitle{Assessing spatial disparities in Internet quality
  using speed tests}.
\newblock  (\bibinfo{year}{2023}).
\newblock


\bibitem[{Cappellari} et~al\mbox{.}(2013)]%
        {Cappellari2013b}
\bibfield{author}{\bibinfo{person}{M. {Cappellari}}, \bibinfo{person}{R.~M.
  {McDermid}}, \bibinfo{person}{K. {Alatalo}}, \bibinfo{person}{L. {Blitz}},
  \bibinfo{person}{M. {Bois}}, \bibinfo{person}{F. {Bournaud}},
  \bibinfo{person}{M. {Bureau}}, \bibinfo{person}{A.~F. {Crocker}},
  \bibinfo{person}{R.~L. {Davies}}, \bibinfo{person}{T.~A. {Davis}},
  \bibinfo{person}{P.~T. {de Zeeuw}}, \bibinfo{person}{P.-A. {Duc}},
  \bibinfo{person}{E. {Emsellem}}, \bibinfo{person}{S. {Khochfar}},
  \bibinfo{person}{D. {Krajnovi{\'c}}}, \bibinfo{person}{H. {Kuntschner}},
  \bibinfo{person}{R. {Morganti}}, \bibinfo{person}{T. {Naab}},
  \bibinfo{person}{T. {Oosterloo}}, \bibinfo{person}{M. {Sarzi}},
  \bibinfo{person}{N. {Scott}}, \bibinfo{person}{P. {Serra}},
  \bibinfo{person}{A.-M. {Weijmans}}, {and} \bibinfo{person}{L.~M. {Young}}.}
  \bibinfo{year}{2013}\natexlab{}.
\newblock \showarticletitle{{The ATLAS$^{3D}$ project - XX. Mass-size and
  mass-{$\sigma$} distributions of early-type galaxies: bulge fraction drives
  kinematics, mass-to-light ratio, molecular gas fraction and stellar initial
  mass function}}.
\newblock \bibinfo{journal}{\emph{MNRAS}}  \bibinfo{volume}{432}
  (\bibinfo{year}{2013}), \bibinfo{pages}{1862--1893}.
\newblock
\urldef\tempurl%
\url{https://doi.org/10.1093/mnras/stt644}
\showDOI{\tempurl}
\showeprint{1208.3523}


\bibitem[Clark and Wedeman(2021)]%
        {clark2021measurement}
\bibfield{author}{\bibinfo{person}{David~D Clark} {and} \bibinfo{person}{Sara
  Wedeman}.} \bibinfo{year}{2021}\natexlab{}.
\newblock \showarticletitle{Measurement, Meaning and Purpose: Exploring the
  M-Lab NDT Dataset}. In \bibinfo{booktitle}{\emph{TPRC49: The 49th Research
  Conference on Communication, Information and Internet Policy}}.
\newblock


\bibitem[Coalition(2022)]%
        {broadband_mapping_coalition_2022}
\bibfield{author}{\bibinfo{person}{Broadband~Mapping Coalition}.}
  \bibinfo{year}{2022}\natexlab{}.
\newblock \bibinfo{title}{Assessing Available Databases for Network Performance
  Measurements: Speed Test Datasets}.
\newblock
  \bibinfo{howpublished}{\url{https://broadbandmappingcoalition.org/assessing-available-databases-network-performance-measurements-speed-test-datasets/}}.
\newblock
\newblock
\shownote{Accessed: 2024-07-31}.


\bibitem[Cressie(1988)]%
        {cressie1988spatial}
\bibfield{author}{\bibinfo{person}{Noel Cressie}.}
  \bibinfo{year}{1988}\natexlab{}.
\newblock \showarticletitle{Spatial prediction and ordinary kriging}.
\newblock \bibinfo{journal}{\emph{Mathematical geology}}  \bibinfo{volume}{20}
  (\bibinfo{year}{1988}), \bibinfo{pages}{405--421}.
\newblock


\bibitem[Duque et~al\mbox{.}(2012)]%
        {duque2012max}
\bibfield{author}{\bibinfo{person}{Juan~C Duque}, \bibinfo{person}{Luc
  Anselin}, {and} \bibinfo{person}{Sergio~J Rey}.}
  \bibinfo{year}{2012}\natexlab{}.
\newblock \showarticletitle{The max-p-regions problem}.
\newblock \bibinfo{journal}{\emph{Journal of Regional Science}}
  \bibinfo{volume}{52}, \bibinfo{number}{3} (\bibinfo{year}{2012}),
  \bibinfo{pages}{397--419}.
\newblock


\bibitem[Efron and Tibshirani(1994)]%
        {efron1994introduction}
\bibfield{author}{\bibinfo{person}{Bradley Efron} {and}
  \bibinfo{person}{Robert~J Tibshirani}.} \bibinfo{year}{1994}\natexlab{}.
\newblock \showarticletitle{An introduction to the bootstrap: CRC press}.
\newblock \bibinfo{journal}{\emph{Ekman, P., \& Friesen, WV (1978). Manual for
  the facial action coding system}} (\bibinfo{year}{1994}).
\newblock


\bibitem[Engineering(2024)]%
        {uber_h3}
\bibfield{author}{\bibinfo{person}{Uber Engineering}.}
  \bibinfo{year}{2024}\natexlab{}.
\newblock \bibinfo{title}{Introducing H3: Uber's Hexagonal Hierarchical Spatial
  Index}.
\newblock \bibinfo{howpublished}{\url{https://www.uber.com/blog/h3/}}.
\newblock
\newblock
\shownote{Accessed: 2024-08-07}.


\bibitem[Feamster and Livingood(2020)]%
        {feamster2020measuring}
\bibfield{author}{\bibinfo{person}{Nick Feamster} {and} \bibinfo{person}{Jason
  Livingood}.} \bibinfo{year}{2020}\natexlab{}.
\newblock \showarticletitle{Measuring internet speed: current challenges and
  future recommendations}.
\newblock \bibinfo{journal}{\emph{Commun. ACM}} \bibinfo{volume}{63},
  \bibinfo{number}{12} (\bibinfo{year}{2020}), \bibinfo{pages}{72--80}.
\newblock


\bibitem[{Federal Communications Commission}(2022)]%
        {FCC2022}
\bibfield{author}{\bibinfo{person}{{Federal Communications Commission}}.}
  \bibinfo{year}{2022}\natexlab{}.
\newblock \bibinfo{title}{{FTC Takes Action Against Frontier for Lying about
  Internet Speeds and Ripping Off Customers Who Paid High-Speed Prices for Slow
  Service}}.
\newblock \bibinfo{howpublished}{Press Release}.
\newblock
\newblock
\shownote{\url{https://www.ftc.gov/news-events/news/pressreleases/2022/05/ftc-takes-action-against-frontier-lying-about-internet-speeds-ripping-customers-who-paid-highspeed}}.


\bibitem[{Federal Communications Commission (FCC)}(2022)]%
        {NBM}
\bibfield{author}{\bibinfo{person}{{Federal Communications Commission (FCC)}}.}
  \bibinfo{year}{2022}\natexlab{}.
\newblock \bibinfo{booktitle}{\emph{{{FCC National Broadband Map}}}}.
\newblock {FCC National Broadband Map}.
\newblock
\urldef\tempurl%
\url{https://broadbandmap.fcc.gov}
\showURL{%
\tempurl}


\bibitem[Fiduccia(2022)]%
        {fiduccia2022deconstructing}
\bibfield{author}{\bibinfo{person}{Peter~Cody Fiduccia}.}
  \bibinfo{year}{2022}\natexlab{}.
\newblock \bibinfo{booktitle}{\emph{Deconstructing the Digital Divide: The
  Geography, Demography, and Spatial Dependence of Internet Stability in the
  US}}.
\newblock \bibinfo{publisher}{Cornell University}.
\newblock


\bibitem[Ge et~al\mbox{.}(2001)]%
        {ge2001hierarchical}
\bibfield{author}{\bibinfo{person}{Zihui Ge}, \bibinfo{person}{Daniel~R
  Figueiredo}, \bibinfo{person}{Sharad Jaiswal}, {and} \bibinfo{person}{Lixin
  Gao}.} \bibinfo{year}{2001}\natexlab{}.
\newblock \showarticletitle{Hierarchical structure of the logical Internet
  graph}. In \bibinfo{booktitle}{\emph{Scalability and Traffic Control in IP
  Networks}}, Vol.~\bibinfo{volume}{4526}. SPIE, \bibinfo{pages}{208--222}.
\newblock


\bibitem[Hasan et~al\mbox{.}(2014)]%
        {hasan2014trade}
\bibfield{author}{\bibinfo{person}{Syed Hasan}, \bibinfo{person}{Sergey
  Gorinsky}, \bibinfo{person}{Constantine Dovrolis}, {and}
  \bibinfo{person}{Ramesh~K Sitaraman}.} \bibinfo{year}{2014}\natexlab{}.
\newblock \showarticletitle{Trade-offs in optimizing the cache deployments of
  CDNs}. In \bibinfo{booktitle}{\emph{IEEE INFOCOM 2014-IEEE conference on
  computer communications}}. IEEE, \bibinfo{pages}{460--468}.
\newblock


\bibitem[Hastie et~al\mbox{.}(2009)]%
        {hastie2009elements}
\bibfield{author}{\bibinfo{person}{Trevor Hastie}, \bibinfo{person}{Robert
  Tibshirani}, \bibinfo{person}{Jerome~H Friedman}, {and}
  \bibinfo{person}{Jerome~H Friedman}.} \bibinfo{year}{2009}\natexlab{}.
\newblock \bibinfo{booktitle}{\emph{The elements of statistical learning: data
  mining, inference, and prediction}}. Vol.~\bibinfo{volume}{2}.
\newblock \bibinfo{publisher}{Springer}.
\newblock


\bibitem[Huang et~al\mbox{.}(2008)]%
        {huang2008measuring}
\bibfield{author}{\bibinfo{person}{Cheng Huang}, \bibinfo{person}{Angela Wang},
  \bibinfo{person}{Jin Li}, {and} \bibinfo{person}{Keith~W Ross}.}
  \bibinfo{year}{2008}\natexlab{}.
\newblock \showarticletitle{Measuring and evaluating large-scale CDNs}. In
  \bibinfo{booktitle}{\emph{ACM IMC}}, Vol.~\bibinfo{volume}{8}.
  \bibinfo{pages}{15--29}.
\newblock


\bibitem[Janc et~al\mbox{.}(2022)]%
        {janc2022spatial}
\bibfield{author}{\bibinfo{person}{Krzysztof Janc}, \bibinfo{person}{Dariusz
  Ilnicki}, {and} \bibinfo{person}{Wojciech Jurkowski}.}
  \bibinfo{year}{2022}\natexlab{}.
\newblock \showarticletitle{Spatial regularities in Internet performance at a
  local scale: The case of Poland}.
\newblock \bibinfo{journal}{\emph{Moravian Geographical Reports}}
  \bibinfo{volume}{30}, \bibinfo{number}{3} (\bibinfo{year}{2022}),
  \bibinfo{pages}{163--178}.
\newblock


\bibitem[Jiang et~al\mbox{.}(2023)]%
        {jiang2023mobile}
\bibfield{author}{\bibinfo{person}{Hanyang Jiang},
  \bibinfo{person}{Henry~Shaowu Yuchi}, \bibinfo{person}{Elizabeth Belding},
  \bibinfo{person}{Ellen Zegura}, {and} \bibinfo{person}{Yao Xie}.}
  \bibinfo{year}{2023}\natexlab{}.
\newblock \showarticletitle{Mobile Internet Quality Estimation using
  Self-Tuning Kernel Regression}.
\newblock \bibinfo{journal}{\emph{arXiv preprint arXiv:2311.05641}}
  (\bibinfo{year}{2023}).
\newblock


\bibitem[Jordahl et~al\mbox{.}(2020)]%
        {geopandas}
\bibfield{author}{\bibinfo{person}{Kelsey Jordahl}, \bibinfo{person}{Joris~Van
  den Bossche}, \bibinfo{person}{Martin Fleischmann}, \bibinfo{person}{Jacob
  Wasserman}, \bibinfo{person}{James McBride}, \bibinfo{person}{Jeffrey
  Gerard}, \bibinfo{person}{Jeff Tratner}, \bibinfo{person}{Matthew Perry},
  \bibinfo{person}{Adrian~Garcia Badaracco}, \bibinfo{person}{Carson Farmer},
  \bibinfo{person}{Geir~Arne Hjelle}, \bibinfo{person}{Alan~D. Snow},
  \bibinfo{person}{Micah Cochran}, \bibinfo{person}{Sean Gillies},
  \bibinfo{person}{Lucas Culbertson}, \bibinfo{person}{Matt Bartos},
  \bibinfo{person}{Nick Eubank}, \bibinfo{person}{maxalbert},
  \bibinfo{person}{Aleksey Bilogur}, \bibinfo{person}{Sergio Rey},
  \bibinfo{person}{Christopher Ren}, \bibinfo{person}{Dani Arribas-Bel},
  \bibinfo{person}{Leah Wasser}, \bibinfo{person}{Levi~John Wolf},
  \bibinfo{person}{Martin Journois}, \bibinfo{person}{Joshua Wilson},
  \bibinfo{person}{Adam Greenhall}, \bibinfo{person}{Chris Holdgraf},
  \bibinfo{person}{Filipe}, {and} \bibinfo{person}{François Leblanc}.}
  \bibinfo{year}{2020}\natexlab{}.
\newblock \bibinfo{booktitle}{\emph{geopandas/geopandas: v0.8.1}}.
\newblock
\urldef\tempurl%
\url{https://doi.org/10.5281/zenodo.3946761}
\showDOI{\tempurl}


\bibitem[K{\"a}rkk{\"a}inen and Fr{\"a}nti(2000)]%
        {karkkainen2000minimization}
\bibfield{author}{\bibinfo{person}{Ismo K{\"a}rkk{\"a}inen} {and}
  \bibinfo{person}{Pasi Fr{\"a}nti}.} \bibinfo{year}{2000}\natexlab{}.
\newblock \showarticletitle{Minimization of the value of Davies-Bouldin index}.
  In \bibinfo{booktitle}{\emph{Proceedings of the IASTED International
  Conference on Signal Processing and Communications (SPC’2000). IASTED/ACTA
  Press}}. \bibinfo{pages}{426--432}.
\newblock


\bibitem[Kaufman(2013)]%
        {kaufman2013chicago}
\bibfield{author}{\bibinfo{person}{Jerome~L Kaufman}.}
  \bibinfo{year}{2013}\natexlab{}.
\newblock \showarticletitle{Chicago: segregation and the new urban poverty}.
\newblock In \bibinfo{booktitle}{\emph{Urban segregation and the welfare
  state}}. \bibinfo{publisher}{Routledge}, \bibinfo{pages}{45--63}.
\newblock


\bibitem[Lab(2024)]%
        {mlab}
\bibfield{author}{\bibinfo{person}{Measurement Lab}.}
  \bibinfo{year}{2024}\natexlab{}.
\newblock \bibinfo{title}{{MLab Test Your Speed}}.
\newblock \bibinfo{howpublished}{\url{https://speed.measurementlab.net/}}.
\newblock
\newblock
\shownote{Accessed: 2024-08-07}.


\bibitem[Lee et~al\mbox{.}(2023)]%
        {lee2023analyzing}
\bibfield{author}{\bibinfo{person}{Hyeongseong Lee}, \bibinfo{person}{Udit
  Paul}, \bibinfo{person}{Arpit Gupta}, \bibinfo{person}{Elizabeth Belding},
  {and} \bibinfo{person}{Mengyang Gu}.} \bibinfo{year}{2023}\natexlab{}.
\newblock \showarticletitle{Analyzing Disparity and Temporal Progression of
  Internet Quality through Crowdsourced Measurements with Bias-Correction}.
\newblock \bibinfo{journal}{\emph{arXiv preprint arXiv:2310.16136}}
  (\bibinfo{year}{2023}).
\newblock


\bibitem[MacMillan et~al\mbox{.}(2023)]%
        {macmillan2023comparative}
\bibfield{author}{\bibinfo{person}{Kyle MacMillan}, \bibinfo{person}{Tarun
  Mangla}, \bibinfo{person}{James Saxon}, \bibinfo{person}{Nicole~P Marwell},
  {and} \bibinfo{person}{Nick Feamster}.} \bibinfo{year}{2023}\natexlab{}.
\newblock \showarticletitle{A comparative analysis of ookla speedtest and
  measurement labs network diagnostic test (ndt7)}.
\newblock \bibinfo{journal}{\emph{Proceedings of the ACM on Measurement and
  Analysis of Computing Systems}} \bibinfo{volume}{7}, \bibinfo{number}{1}
  (\bibinfo{year}{2023}), \bibinfo{pages}{1--26}.
\newblock


\bibitem[Marques et~al\mbox{.}(2024)]%
        {marques2024we}
\bibfield{author}{\bibinfo{person}{Jonatas Marques}, \bibinfo{person}{Alexis
  Schrubbe}, \bibinfo{person}{Nicole~P Marwell}, {and} \bibinfo{person}{Nick
  Feamster}.} \bibinfo{year}{2024}\natexlab{}.
\newblock \showarticletitle{Are We Up to the Challenge? An analysis of the FCC
  Broadband Data Collection Fixed Internet Availability Challenges}.
\newblock \bibinfo{journal}{\emph{arXiv preprint arXiv:2404.04189}}
  (\bibinfo{year}{2024}).
\newblock


\bibitem[Martin and Dogar(2023)]%
        {martin2023divided}
\bibfield{author}{\bibinfo{person}{Noah Martin} {and} \bibinfo{person}{Fahad
  Dogar}.} \bibinfo{year}{2023}\natexlab{}.
\newblock \showarticletitle{Divided at the Edge-Measuring Performance and the
  Digital Divide of Cloud Edge Data Centers}.
\newblock \bibinfo{journal}{\emph{Proceedings of the ACM on Networking}}
  \bibinfo{volume}{1}, \bibinfo{number}{CoNEXT3} (\bibinfo{year}{2023}),
  \bibinfo{pages}{1--23}.
\newblock


\bibitem[Moran(1950)]%
        {moran1950notes}
\bibfield{author}{\bibinfo{person}{P.~A.~P. Moran}.}
  \bibinfo{year}{1950}\natexlab{}.
\newblock \showarticletitle{Notes on continuous stochastic phenomena}.
\newblock \bibinfo{journal}{\emph{Biometrika}} \bibinfo{volume}{37},
  \bibinfo{number}{1/2} (\bibinfo{year}{1950}), \bibinfo{pages}{17--23}.
\newblock


\bibitem[{New York State Office of the Attorney General}(2020)]%
        {NYAG2020}
\bibfield{author}{\bibinfo{person}{{New York State Office of the Attorney
  General}}.} \bibinfo{year}{2020}\natexlab{}.
\newblock \bibinfo{booktitle}{\emph{{New York Internet Health Test}}}.
\newblock
\urldef\tempurl%
\url{https://ag.ny.gov/SpeedTest}
\showURL{%
\tempurl}


\bibitem[Ookla(2024)]%
        {ookla}
\bibfield{author}{\bibinfo{person}{Ookla}.} \bibinfo{year}{2024}\natexlab{}.
\newblock \bibinfo{title}{Ookla Speedtest}.
\newblock \bibinfo{howpublished}{\url{https://www.speedtest.net/}}.
\newblock
\newblock
\shownote{Accessed: 2024-08-07}.


\bibitem[Paul et~al\mbox{.}(2023)]%
        {paul2023decoding}
\bibfield{author}{\bibinfo{person}{Udit Paul}, \bibinfo{person}{Vinothini
  Gunasekaran}, \bibinfo{person}{Jiamo Liu}, \bibinfo{person}{Tejas~N
  Narechania}, \bibinfo{person}{Arpit Gupta}, {and} \bibinfo{person}{Elizabeth
  Belding}.} \bibinfo{year}{2023}\natexlab{}.
\newblock \showarticletitle{Decoding the Divide: Analyzing Disparities in
  Broadband Plans Offered by Major US ISPs}. In
  \bibinfo{booktitle}{\emph{Proceedings of the ACM SIGCOMM 2023 Conference}}.
  \bibinfo{pages}{578--591}.
\newblock


\bibitem[Paul et~al\mbox{.}(2021)]%
        {paul2021characterizing}
\bibfield{author}{\bibinfo{person}{Udit Paul}, \bibinfo{person}{Jiamo Liu},
  \bibinfo{person}{Vivek Adarsh}, \bibinfo{person}{Mengyang Gu},
  \bibinfo{person}{Arpit Gupta}, {and} \bibinfo{person}{Elizabeth Belding}.}
  \bibinfo{year}{2021}\natexlab{}.
\newblock \showarticletitle{Characterizing performance inequity across us ookla
  speedtest users}.
\newblock \bibinfo{journal}{\emph{arXiv preprint arXiv:2110.12038}}
  (\bibinfo{year}{2021}).
\newblock


\bibitem[Paul et~al\mbox{.}(2022a)]%
        {paul2022characterizing}
\bibfield{author}{\bibinfo{person}{Udit Paul}, \bibinfo{person}{Jiamo Liu},
  \bibinfo{person}{David Farias-Llerenas}, \bibinfo{person}{Vivek Adarsh},
  \bibinfo{person}{Arpit Gupta}, {and} \bibinfo{person}{Elizabeth Belding}.}
  \bibinfo{year}{2022}\natexlab{a}.
\newblock \showarticletitle{Characterizing internet access and quality
  inequities in california m-lab measurements}. In
  \bibinfo{booktitle}{\emph{Proceedings of the 5th ACM SIGCAS/SIGCHI Conference
  on Computing and Sustainable Societies}}. \bibinfo{pages}{257--265}.
\newblock


\bibitem[Paul et~al\mbox{.}(2022b)]%
        {paul2022importance}
\bibfield{author}{\bibinfo{person}{Udit Paul}, \bibinfo{person}{Jiamo Liu},
  \bibinfo{person}{Mengyang Gu}, \bibinfo{person}{Arpit Gupta}, {and}
  \bibinfo{person}{Elizabeth Belding}.} \bibinfo{year}{2022}\natexlab{b}.
\newblock \showarticletitle{The importance of contextualization of crowdsourced
  active speed test measurements}. In \bibinfo{booktitle}{\emph{Proceedings of
  the 22nd ACM Internet Measurement Conference}}. \bibinfo{pages}{274--289}.
\newblock


\bibitem[Pedregosa et~al\mbox{.}(2011)]%
        {scikit-learn}
\bibfield{author}{\bibinfo{person}{Fabian Pedregosa}, \bibinfo{person}{Ga{\"e}l
  Varoquaux}, \bibinfo{person}{Alexandre Gramfort}, \bibinfo{person}{Vincent
  Michel}, \bibinfo{person}{Bertrand Thirion}, \bibinfo{person}{Olivier
  Grisel}, \bibinfo{person}{Mathieu Blondel}, \bibinfo{person}{Peter
  Prettenhofer}, \bibinfo{person}{Ron Weiss}, \bibinfo{person}{Vincent
  Dubourg}, {et~al\mbox{.}}} \bibinfo{year}{2011}\natexlab{}.
\newblock \showarticletitle{Scikit-learn: Machine learning in Python}.
\newblock \bibinfo{journal}{\emph{Journal of machine learning research}}
  \bibinfo{volume}{12}, \bibinfo{number}{Oct} (\bibinfo{year}{2011}),
  \bibinfo{pages}{2825--2830}.
\newblock


\bibitem[{Pennsylvania State University and Measurement Lab}(2019)]%
        {PennStateMeasurementLab2019}
\bibfield{author}{\bibinfo{person}{{Pennsylvania State University and
  Measurement Lab}}.} \bibinfo{year}{2019}\natexlab{}.
\newblock \bibinfo{booktitle}{\emph{{Broadband Availability and Access in Rural
  Pennsylvania}}}.
\newblock
\urldef\tempurl%
\url{https://www.rural.pa.gov/publications/broadband.cfm}
\showURL{%
\tempurl}


\bibitem[{QGIS Development Team}(2009)]%
        {qgis}
\bibfield{author}{\bibinfo{person}{{QGIS Development Team}}.}
  \bibinfo{year}{2009}\natexlab{}.
\newblock \bibinfo{booktitle}{\emph{QGIS Geographic Information System}}.
\newblock Open Source Geospatial Foundation.
\newblock
\urldef\tempurl%
\url{http://qgis.org}
\showURL{%
\tempurl}


\bibitem[Radovanov and Marciki{\'c}(2014)]%
        {radovanov2014comparison}
\bibfield{author}{\bibinfo{person}{Boris Radovanov} {and}
  \bibinfo{person}{Aleksandra Marciki{\'c}}.} \bibinfo{year}{2014}\natexlab{}.
\newblock \showarticletitle{A comparison of four different block bootstrap
  methods}.
\newblock \bibinfo{journal}{\emph{Croatian Operational Research Review}}
  (\bibinfo{year}{2014}), \bibinfo{pages}{189--202}.
\newblock


\bibitem[Redlands(2011)]%
        {arcgis}
\bibfield{author}{\bibinfo{person}{CA: Environmental Systems Research~Institute
  Redlands}.} \bibinfo{year}{2011}\natexlab{}.
\newblock \bibinfo{title}{ArcGIS Desktop: Release 10}.
\newblock
\newblock


\bibitem[Rigol et~al\mbox{.}(2001)]%
        {rigol2001artificial}
\bibfield{author}{\bibinfo{person}{Juan~P Rigol}, \bibinfo{person}{Claire~H
  Jarvis}, {and} \bibinfo{person}{Neil Stuart}.}
  \bibinfo{year}{2001}\natexlab{}.
\newblock \showarticletitle{Artificial neural networks as a tool for spatial
  interpolation}.
\newblock \bibinfo{journal}{\emph{International Journal of Geographical
  Information Science}} \bibinfo{volume}{15}, \bibinfo{number}{4}
  (\bibinfo{year}{2001}), \bibinfo{pages}{323--343}.
\newblock


\bibitem[Saxon and Black(2022)]%
        {saxon2022we}
\bibfield{author}{\bibinfo{person}{James Saxon} {and} \bibinfo{person}{Dan~A
  Black}.} \bibinfo{year}{2022}\natexlab{}.
\newblock \showarticletitle{What we can learn from selected, unmatched data:
  measuring Internet inequality in Chicago}.
\newblock \bibinfo{journal}{\emph{Computers, Environment and Urban Systems}}
  \bibinfo{volume}{98} (\bibinfo{year}{2022}), \bibinfo{pages}{101874}.
\newblock


\bibitem[Sekuli{\'c} et~al\mbox{.}(2020)]%
        {sekulic2020random}
\bibfield{author}{\bibinfo{person}{Aleksandar Sekuli{\'c}},
  \bibinfo{person}{Milan Kilibarda}, \bibinfo{person}{Gerard~BM Heuvelink},
  \bibinfo{person}{Mladen Nikoli{\'c}}, {and} \bibinfo{person}{Branislav
  Bajat}.} \bibinfo{year}{2020}\natexlab{}.
\newblock \showarticletitle{Random forest spatial interpolation}.
\newblock \bibinfo{journal}{\emph{Remote Sensing}} \bibinfo{volume}{12},
  \bibinfo{number}{10} (\bibinfo{year}{2020}), \bibinfo{pages}{1687}.
\newblock


\bibitem[Shahapure and Nicholas(2020)]%
        {shahapure2020cluster}
\bibfield{author}{\bibinfo{person}{Ketan~Rajshekhar Shahapure} {and}
  \bibinfo{person}{Charles Nicholas}.} \bibinfo{year}{2020}\natexlab{}.
\newblock \showarticletitle{Cluster quality analysis using silhouette score}.
  In \bibinfo{booktitle}{\emph{2020 IEEE 7th international conference on data
  science and advanced analytics (DSAA)}}. IEEE, \bibinfo{pages}{747--748}.
\newblock


\bibitem[Sharma et~al\mbox{.}(2022)]%
        {sharma2022benchmarks}
\bibfield{author}{\bibinfo{person}{Ranya Sharma}, \bibinfo{person}{Tarun
  Mangla}, \bibinfo{person}{James Saxon}, \bibinfo{person}{Marc Richardson},
  \bibinfo{person}{Nick Feamster}, {and} \bibinfo{person}{Nicole~P Marwell}.}
  \bibinfo{year}{2022}\natexlab{}.
\newblock \showarticletitle{Benchmarks or Equity? A New Approach to Measuring
  Internet Performance}.
\newblock \bibinfo{journal}{\emph{A New Approach to Measuring Internet
  Performance (August 3, 2022)}} (\bibinfo{year}{2022}).
\newblock


\bibitem[Sharma et~al\mbox{.}(2023b)]%
        {sharma2023measuringprevalencewifibottlenecks}
\bibfield{author}{\bibinfo{person}{Ranya Sharma}, \bibinfo{person}{Marc
  Richardson}, \bibinfo{person}{Guilherme Martins}, {and} \bibinfo{person}{Nick
  Feamster}.} \bibinfo{year}{2023}\natexlab{b}.
\newblock \bibinfo{title}{Measuring the Prevalence of WiFi Bottlenecks in Home
  Access Networks}.
\newblock
\newblock
\showeprint[arxiv]{2311.05499}~[cs.NI]
\urldef\tempurl%
\url{https://arxiv.org/abs/2311.05499}
\showURL{%
\tempurl}


\bibitem[Sharma et~al\mbox{.}(2023a)]%
        {sharma2023first}
\bibfield{author}{\bibinfo{person}{Taveesh Sharma}, \bibinfo{person}{Jonatas
  Marques}, \bibinfo{person}{Nick Feamster}, {and} \bibinfo{person}{Nicole~P
  Marwell}.} \bibinfo{year}{2023}\natexlab{a}.
\newblock \showarticletitle{A First Look at the Spatial and Temporal
  Variability of Internet Performance Data in Hyperlocal Geographies}.
\newblock \bibinfo{journal}{\emph{Available at SSRN 4568668}}
  (\bibinfo{year}{2023}).
\newblock


\bibitem[Sharma et~al\mbox{.}(2024)]%
        {artifacts}
\bibfield{author}{\bibinfo{person}{Taveesh Sharma}, \bibinfo{person}{Paul
  Schmitt}, \bibinfo{person}{Francesco Bronzino}, \bibinfo{person}{Nick
  Feamster}, {and} \bibinfo{person}{Nicole~P Marwell}.}
  \bibinfo{year}{2024}\natexlab{}.
\newblock \bibinfo{title}{Latency Regionalization}.
\newblock
  \bibinfo{howpublished}{\url{https://github.com/noise-lab/latency-regionalization}}.
\newblock
\newblock
\shownote{Accessed: 2024-10-17}.


\bibitem[Shepard(1968)]%
        {shepard1968two}
\bibfield{author}{\bibinfo{person}{Donald Shepard}.}
  \bibinfo{year}{1968}\natexlab{}.
\newblock \showarticletitle{A two-dimensional interpolation function for
  irregularly-spaced data}. In \bibinfo{booktitle}{\emph{Proceedings of the
  1968 23rd ACM national conference}}. \bibinfo{pages}{517--524}.
\newblock


\bibitem[Sommers and Barford(2012)]%
        {sommers2012cell}
\bibfield{author}{\bibinfo{person}{Joel Sommers} {and} \bibinfo{person}{Paul
  Barford}.} \bibinfo{year}{2012}\natexlab{}.
\newblock \showarticletitle{Cell vs. WiFi: On the performance of metro area
  mobile connections}. In \bibinfo{booktitle}{\emph{Proceedings of the 2012
  internet measurement conference}}. \bibinfo{pages}{301--314}.
\newblock


\bibitem[Steinley(2004)]%
        {steinley2004properties}
\bibfield{author}{\bibinfo{person}{Douglas Steinley}.}
  \bibinfo{year}{2004}\natexlab{}.
\newblock \showarticletitle{Properties of the hubert-arable adjusted rand
  index.}
\newblock \bibinfo{journal}{\emph{Psychological methods}} \bibinfo{volume}{9},
  \bibinfo{number}{3} (\bibinfo{year}{2004}), \bibinfo{pages}{386}.
\newblock


\bibitem[Stokes and Thompson(2022)]%
        {natlawreview}
\bibfield{author}{\bibinfo{person}{Sean Stokes} {and}
  \bibinfo{person}{Kathleen~Slattery Thompson}.}
  \bibinfo{year}{2022}\natexlab{}.
\newblock \showarticletitle{With Billions of Dollars of Broadband Funding at
  Stake, the Timing of the Challenge Process to the FCC`s Broadband Map Under
  Increasing Scrutiny}.
\newblock \bibinfo{journal}{\emph{The National Law Review}}
  (\bibinfo{year}{2022}).
\newblock
\urldef\tempurl%
\url{https://natlawreview.com/article/billions-dollars-broadband-funding-stake-timing-challenge-process-to-fcc-s-broadband}
\showURL{%
\tempurl}


\bibitem[Sundaresan et~al\mbox{.}(2014)]%
        {sundaresan2014bismark}
\bibfield{author}{\bibinfo{person}{Srikanth Sundaresan}, \bibinfo{person}{Sam
  Burnett}, \bibinfo{person}{Nick Feamster}, {and} \bibinfo{person}{Walter
  De~Donato}.} \bibinfo{year}{2014}\natexlab{}.
\newblock \showarticletitle{$\{$BISmark$\}$: A testbed for deploying
  measurements and applications in broadband access networks}. In
  \bibinfo{booktitle}{\emph{2014 USENIX Annual Technical Conference (USENIX ATC
  14)}}. \bibinfo{pages}{383--394}.
\newblock


\bibitem[Sundaresan et~al\mbox{.}(2011)]%
        {sundaresan2011broadband}
\bibfield{author}{\bibinfo{person}{Srikanth Sundaresan},
  \bibinfo{person}{Walter De~Donato}, \bibinfo{person}{Nick Feamster},
  \bibinfo{person}{Renata Teixeira}, \bibinfo{person}{Sam Crawford}, {and}
  \bibinfo{person}{Antonio Pescap{\`e}}.} \bibinfo{year}{2011}\natexlab{}.
\newblock \showarticletitle{Broadband internet performance: a view from the
  gateway}.
\newblock \bibinfo{journal}{\emph{ACM SIGCOMM computer communication review}}
  \bibinfo{volume}{41}, \bibinfo{number}{4} (\bibinfo{year}{2011}),
  \bibinfo{pages}{134--145}.
\newblock


\bibitem[Sundaresan et~al\mbox{.}(2017)]%
        {sundaresan2017challenges}
\bibfield{author}{\bibinfo{person}{Srikanth Sundaresan},
  \bibinfo{person}{Xiaohong Deng}, \bibinfo{person}{Yun Feng},
  \bibinfo{person}{Danny Lee}, {and} \bibinfo{person}{Amogh Dhamdhere}.}
  \bibinfo{year}{2017}\natexlab{}.
\newblock \showarticletitle{Challenges in inferring internet congestion using
  throughput measurements}. In \bibinfo{booktitle}{\emph{Proceedings of the
  2017 Internet Measurement Conference}}. \bibinfo{pages}{43--56}.
\newblock


\bibitem[Telecommunications and Administration(2024)]%
        {internetforall2024challenge}
\bibfield{author}{\bibinfo{person}{National Telecommunications} {and}
  \bibinfo{person}{Information Administration}.}
  \bibinfo{year}{2024}\natexlab{}.
\newblock \bibinfo{title}{State and Territory Challenge Process Tracker}.
\newblock
  \bibinfo{howpublished}{\url{https://www.internetforall.gov/state-and-territory-challenge-process-tracker}}.
\newblock
\newblock
\shownote{Accessed: 2024-07-31}.


\end{thebibliography}

\end{document}